\def\submissiondate{April 30, 1997}	

\documentstyle[12pt,twoside,singlespace,amsfonts,mythesis,epsf]{mitthesis}
\input{defs}

\ifsetnotation
\chapter*{Notation}
\hbox to \hsize{\large\sl Symbol \hfill Meaning \hfill First used}
\hrule
\fi	

\defnotationsubx{adag}{a^\dagger}{$a$, $\adag$}{Raising and lowering
operators for normal modes of $H$}
\notation{A}{Arbitrary operator}
\def\Asup#1{A^{(#1)}}
\notationx{Asup}{$\Asup{n}$}{Sequence of operators}
\notationx{Abar}{$\bar A$}{Limit of $\Asup{n}$}

\defnotationsubx{bdag}{b^\dagger}{$b$, $\bdag$}{Raising and lowering
operators for improved vacuum normal modes}

\notationx{Calpha}{$C_\alpha$}{General constraint operator}

\notationx{DDbar}{$\bar D$, $\bar D'$}{Free part of outside
oscillators: initial form}
\notationx{DDprime}{$D$, $D'$}{Free part of outside oscillators: final
form}

\notation{E}{Energy}
\notation{E_0}{Average energy of vacuum-bounded state}

\notationx{fg}{$f$, $g$}{Coefficients of constraints in $\ln\rho$}
\notation{f(x)}{Coupling coefficient ratio}

\notation{g}{Effective number of degrees of freedom at $\Tunk$}
\notation{g(x)}{Continuum eigenfunction}
\defnotation{calg}{{\cal G}}{Space of states of ``ground-state'' modes}

\notationx{Harb}{$H$}{Operator used to define $\calp(E)$}
\notation{H}{Hamiltonian of original vacuum}
\notationx{Hfree}{$\tilde H$}{Space of states of ``free'' modes}
\notationx{altH}{$H'$}{Alternate Hamiltonian with same vacuum}
\notation{H'}{Fictitious Hamiltonian}
\defnotation{calh}{{\cal H}}{Hilbert space of states}
\notationx{calh'}{$\calh'$}{Subspace of $\calh$ on which every
admissible $\rho$ is non-vanishing}

\notation{K}{Potential coupling matrix of vacuum}
\notationx{tildeK}{$\tilde K$}{Potential coupling matrix of reduced vacuum}
\notation{K'}{Potential couplings in $H'$}

\defnotation{lpl}{l_{\text{pl}}}{Planck energy}
\notation{L_1}{Lattice spacing}
\defnotationsub{Lin}{L_\in}{Length of inside region}
\defnotation{Linp}{L_\in\hskip -1.7ex '\hskip 1.0ex}{Equivalent length of system with same S and E}
\def\Linpsup#1{L_\in \hskip -1.7ex '\hskip .3ex^{(#1)}}

\defnotation{mpl}{m_{\text{pl}}}{Planck mass}
\notation{M_0}{Mass of black hole at cutoff}

\defnotationsub{Nfree}{N_\free}{Number of ``free'' modes}
\defnotationsub{Ngs}{N_\gs}{Number of ``ground state'' modes}
\defnotationsub{Nin}{N_\in}{Number of inside oscillators}
\def\Ninn{{N_\in+1}}
\defnotationsub{Nnorm}{N_{\text{norm}}}{Number of ``normal'' modes}
\defnotationsub{Nout}{N_\out}{Number of outside oscillators}
\notationx{Nalpha}{$N_\alpha$, $N_n$}{Normalization factors in normal modes}
\defnotation{caln}{{\cal N}}{Nullspace of $\rho$}

\notationx{Oout}{$O_\out^\alpha$}{Operator made of outside fields}

\notationx{pk}{$p_k(\rho)$}{The $k$th largest eigenvalue of $\rho$}
\defnotationsubx{Px}{{\bf P}_x}{$\bP$ or $\Px$}{Momentum associated with $\bx$}
\defnotationsub{Pu}{{\bf P}_u}{Momentum associated with $\bu$}
\defnotationsub{Pw}{{\bf P}_w}{Momentum associated with $\bw$}
\def\bP{{\bf P}}
\defnotationsub{Py}{{\bf P}_y}{Momentum associated with $\by$}
\defnotationsub{Pz}{{\bf P}_z}{Momentum associated with $\bz$}
\defnotationsub{pop}{{\Bbb P}}{Matrix of pairs of position operators}
\def\popsub{\pop}
\def\popex{\<\pop\>}
\def\popexsub{\popex}
\defnotationsub{pwop}{{\Bbb P}_w}{Matrix of pairs of ${\bf P}_w$ operators}
\def\pwopsub{(\pwop)}

\defnotationsub{pzop}{{\Bbb P}_z}{Matrix of pairs of $\Pz$ operators}

\def\popone{C}			
\def\pzopex{\<\pzop\>}
\def\pzopexsub{\pzopex}
\defnotation{calp}{{\cal P}}{Space of admissible density matrices}
\defnotation{calpmanyeq}{\calp_=(E;V_1, V_2, \ldots)}{Space of $\rho$ 
satisfying equality constraints}
\defnotation{calpmanyle}{\calp_\le(E;V_1, V_2, \ldots)}{Space of $\rho$
satisfying bounds}

\notationx{Qgrand}{$Q$}{Grand partition function}
\notation{Q}{Transformation between $\bx$ and $\bw$}
\notationx{Qout}{$Q_{\text{out}}^a$}{Operator quadratic in outside fields}

\notationx{Rradius}{$R$}{Radius of inside region}
\notation{R}{Transformation between $\bx_\out$ and $\bw_\out$}

\notation{s(p)}{Continuous version of $-p\ln p$}

\defnotation{Sth}{S_{\text{th}}}{Entropy of a system in the
thermal state}

\notation{T}{Temperature of vacuum-bounded state}
\defnotation{Tunk}{T_{\text{unk}}}{Temperature of unknown physics}
\notationx{tildeT}{$\tilde T$}{Kinetic couplings in reduced vacuum}
\notation{T'}{Kinetic couplings in $H'$}

\defnotationsub{bu}{{\bf u}}{Normal-mode coordinates for reduced problem}
\notation{U}{Normal modes of fictitious Hamiltonian in terms of $\bf
w$ coordinates}
\def\uvacinv{{U^\vac}^{\raise .5ex\hbox{$\scriptstyle -1$}}}

\notationx{Valpha}{$V_\alpha$}{Generalized constraint value}

\defnotationsub{bw}{{\bf w}}{Oscillators adapted to reduced problem}
\defnotationsub{walpha}{{\bf w}^\alpha}{Eigenvector of reduced Hamiltonian}
\notation{W}{Transformation between $Z$ and $Y$}
\defnotationsub{wop}{{\Bbb W}}{Matrix of pairs of ${\bf w}$ operators}
\def\wopsub{\wop}

\def\bx{{\bf x}}
\notationx{bx}{$\bx$}{Oscillator coordinates}
\defnotationsub{xop}{{\Bbb X}}{Matrix of pairs of position operators}
\def\xopsub{\xop}
\def\xopex{\<\xop\>}
\def\xopexsub{\xopex}

\defnotationsub{by}{{\bf y}}{Improved normal-mode coordinates for the vacuum}
\notation{Y}{Matrix of improved vacuum normal modes}

\defnotationsub{bz}{{\bf z}}{Normal-mode coordinates that diagonalize $H$}
\notation{Z}{Matrix of normal modes of $H$}
\defnotationsub{zop}{{\Bbb Z}}{Matrix of pairs of $\bf z$ operators}

\def\zopex{\<\zop\>}
\def\zopexsub{\zopex}

\vspace{2ex}

\notationx{beta}{$\beta$}{Inverse temperature}

\notationx{gamma}{$\gamma$}{Euler's constant}

\defnotation{deltaabnn}{\delta_{\text{abn}}^{(n)}}{Contribution of
abnormal mode to mode sum}
\notationx{d(E)}{$\ddelta A(E)$}{Difference in quantity $A$ at same energy}
\notationx{d(T)}{$\ddelta A(T)$}{Difference in quantity $A$ at same temperature}
\notationx{Delta}{$\Delta$}{Fraction by which equivalent rigid system is larger}

\notationx{pi(x)}{$\pi(x)$}{Scalar field momentum}

\notationx{rho}{$\rho$}{Density matrix}
\def\rhosup#1{\rho^{(#1)}}
\notationx{rhosup}{$\rhosup{n}$}{Sequence of density matrices}
\notationx{rhosuphat}{$\hat\rhosup{n}$}{Cauchy subsequence of density
matrix sequence}
\defnotationsub{rhobar}{\bar\rho}{Limit of a sequence of density matrices}
\def\rhofree{\tilde\rho}
\def\drhofree{\widetilde{\ddelta\rho}}

\notationx{tau}{$\tau$}{Temperature scaled by $\Lin$}
\notationx{tau'}{$\tau'$}{Temperature scaled by $\Linp$}

\notationx{phi(x)}{$\phi(x)$}{Scalar field value}

\notationx{digamma}{$\psi(x)$}{Digamma function}

\notationx{omega}{$\omega_\alpha$}{Normal mode frequencies}
\notationx{omega0}{$\omega^{(0)}_\alpha$}{Normal mode frequencies in original vacuum}
\notationx{Omega}{$\Omega$}{Diagonal matrix of frequencies} 

\vspace{2ex}

\def\thenorm{\|\!\cdot\!\|}
\notationx{norm}{$\thenorm$}{Norm of a state in $\calh$}
\notationx{opnorm}{$\thenorm$}{Bound on a bounded operator}
\notationx{hsnorm}{$\thenorm_2$}{Hilbert-Schmidt norm}

\notationx{downn}{$\downarrow_N$}{Submatrix with both indices $\le N$}
\def\downn#1{\hbox{$\downarrow_#1$}\notationlabel{downn}}

\notationx{upn}{$\uparrow^N$}{Submatrix with both indices $> N$}
\def\upn#1{\hbox{$\uparrow^#1$}\notationlabel{upn}}

\notationx{posdef}{$O \ge 0$}{Operator is positive semidefinite}
\notationx{tilde}{$\widetilde O$}{Operator acting only on ``free'' modes}

\defnotationgrpx{free}{\protect\text{free}}{$X_{\protect\text{free}}$}{Components
that can be excited in a vacuum-bounded state} 
\defnotationgrpx{gs}{\text{gs}}{$X_{\protect\text{gs}}$}{Components that
must be in their ground state}
\defnotationgrpx{in}{\protect\text{in}}{$X_{\protect\text{in}}$}{Inside
components}
\defnotationgrpx{mid}{\text{mid}}{$X_{\protect\text{mid}}$}{Components
$N_{\text{in}}$ through $N_{\text{free}}$}
\defnotationgrpx{out}{\text{out}}{$X_{\protect\text{out}}$}{Outside
components}
\defnotationgrpx{rb}{\text{rb}}{$X^{\protect\text{rb}}$}{Quantities in
rigid box problem}
\defnotationgrpx{vac}{\text{vac}}{$X^{\protect\text{vac}}$}{Vacuum quantities}
\begin{document}



\title{Vacuum-Bounded States and the Entropy of \\ Black Hole Evaporation}
\author{Ken D. Olum}
\prevdegrees{B.S. Mathematics, Stanford University, 1982}
\department{Department of Physics}
\degree{Doctor of Philosophy}
\degreemonth{June}
\degreeyear{1997}
\thesisdate{\submissiondate}
\supervisor{Alan H. Guth}{Weisskopf Professor of Physics}
\chairman{George F. Koster}{Chairman, Physics Graduate Committee}
\maketitle
\newpage

\mbox{}\thispagestyle{empty}\newpage


\setnotationtrue

\thispagestyle{empty}
\begin{abstractpage}

We call a state ``vacuum bounded'' if every measurement performed
outside a specified interior region gives the same result as in the
vacuum.  We compute the maximum entropy of a vacuum-bounded state with
a given energy for a one-dimensional model, with the aid of numerical
calculations on a lattice.  For large energies we show that a
vacuum-bounded system with length $\Lin$ and a given energy has
entropy no more than $S^\rb + {1\over 6} \ln S^\rb$, where $S^\rb$ is
the entropy in a rigid box with the same size and energy.  Assuming
that the state resulting from the evaporation of a black hole is
similar to a vacuum-bounded state, and that the similarity between
vacuum-bounded and rigid box problems extends from 1 to 3 dimensions, we
apply these results to the black hole information paradox.  Under
these assumptions we conclude that large amounts of information cannot
be emitted in the final explosion of a black hole.

We also consider vacuum-bounded states at very low energies and come
to the surprising conclusion that the entropy of such a state can be
much higher than that of a rigid box state with the same energy.  For
a fixed $E$ we let $\Linp$ be the length of a rigid box which gives the
same entropy as a vacuum-bounded state of length $\Lin$.  In the
$E\rightarrow 0$ limit we conjecture that the ratio $\Linp/\Lin$ grows
without bound and support this conjecture with numerical computations.

\end{abstractpage}
\newpage

\thispagestyle{empty}
\vspace*{1.5in}
\begin{centering}
To Judy and Valerie, with love.\\
\end{centering}
\newpage


\tableofcontents
\markboth{Contents}{Contents}
\newpage
\listoffigures
\newpage

\newpage

\thispagestyle{plain}
\section*{Acknowledgments}

The author would like to thank Alan Guth and Sean Carroll for much
advice and assistance; Niklas Dellby, Edward Farhi, Dan Freedman, Ken
Halpern, Andy Latto, Leonid Levitov, Arthur Lue, Keith Ramsay and
Peter Unrau for helpful conversations; Alan Guth, Edward Farhi and
Leonid Levitov for serving on the thesis committee; Judy Anderson and
Valerie White for proofreading; Harlequin, Inc.\ for providing a copy
of their LispWorks\trademark\ product on which some of the
computations were done; Bruno Haible and Marcus Daniels for their
CLISP Common Lisp implementation; and Kevin Broughan and William Press
for making available a version of {\em Numerical Recipes\/} translated
into LISP\footnote{These routines and many others are now available on
a CDROM \protect\cite{numrecip:cdrom}}.

Portions of this thesis are reprinted from
\cite{olum:paper}, copyright 1997 The American Physical Society, with
permission.

This work was supported in part by funds provided by the
U.S. Department of Energy (D.O.E.) under cooperative research
agreement DE-FC02-94ER40818.

\newpage

\setnotationfalse		

\chapter{Introduction}

\section{Background}

Since the discovery of black hole radiation by Hawking
\cite{hawking:orig}, the fate of information falling into a black hole
has been a mystery.  (See
\cite{page:review,preskill:review,banks:review} for reviews.)  If
Hawking's semiclassical calculation is correct, then the outgoing
radiation is purely thermal and the outgoing photons are uncorrelated
to each other and to the matter which formed the black hole.  If the
evaporation is complete, and if the thermal nature of the radiation
persists throughout the evaporation, then the original information is
lost.  Thus if the black hole is formed from a quantum-mechanically
pure state, there will nevertheless be a mixed state after the
evaporation is complete.  This is the position held by Hawking
(e.g.\ see \cite{hawking:virtual}), but it violates CPT and may lead to
difficulties with energy conservation and causality
\cite{page:review,giddings:infoloss,banks:pure-mixed,%
strominger:unitary}.

If information is not lost in black hole evaporation, there are
several possibilities. One is that the black hole does not evaporate
completely, but instead produces one or more Planck-scale remnants
(e.g.\ see \cite{banks:review}).  Another possibility is that the
information disappears into a baby universe \cite{strominger:baby}.
In this scenario the quantum-mechanical pure state is preserved, but
parts of it are inaccessible to observation.  It is also possible that
the radiation is not really thermal, even at early times, because of a
complementarity principle 
\cite{hooft:early,hooft:interp,susskind:strings,verlinde:complement}
or the inapplicability of the semiclassical approach
\cite{esko:nosemiclass,bose:nosemiclass,casher:nosemiclass,stephens:noinfoloss},
and thus that the information is encoded in subtle correlations in the
radiation.  In this case the black hole could act like a normal object
with the entropy describing internal degrees of freedom.  Some results
from string theory
\cite{strominger:string,esko:string,horowitz:review,maldacena:thesis} tend to
confirm this view.

Even if the radiation is thermal and uncorrelated during most of the
evaporation, there is no reason to believe that it remains thermal
near the endpoint of the evaporation.  The late-time radiation is
presumably governed by an unknown theory of quantum gravity, and may
well have correlations to the radiation emitted earlier.\footnote{This
is conceivable because at early times information in the outgoing
radiation can be correlated with information in the ingoing
negative-energy flux.}  However, it is generally believed that
late-time radiation cannot resolve the information paradox
\cite{aharonov:orig,preskill:review,banks:review}.  The argument goes
as follows: While the black hole is large, it is presumably radiating
high-entropy thermal radiation.  If the final explosion is to restore
a pure state, it must radiate as much entropy\footnote{Here and
throughout this thesis, ``entropy'' means fine-grained
quantum-mechanical entropy.}  as was radiated in earlier times.
However, by the time the black hole reaches the point where unknown
physics could come into play, there is little energy remaining.  To
radiate a lot of information with little energy requires a long period
of time, and thus the ``final explosion'' looks more like a long-lived
remnant.

However, Wilczek and Holzhey \cite{wilczek:mirror} argue from a moving mirror
model that a state with high entropy can nevertheless be purified
with arbitrarily low energy cost.  In certain ways, their model looks
more like a remnant theory than a complete-evaporation theory, but it
still appears to cast some doubt on the standard argument above.

In any case, this argument requires bounding the entropy that can be
contained in a particular region with a fixed energy.  In the case of
a region with reflecting walls, this is the question of finding the
thermal state of quantum fields in a box.  For a spherical box and a
particular field theory the problem is easily solved.  But with a
region of complex shape, or where one wishes to make a statement
intended to apply to all field theories, the situation is more
complicated.  Bekenstein \cite{bekenstein:talk,bekenstein:argue}
argues that such a universal bound exists, but Unruh and Wald
\cite{unruh-wald:argue} disagree.

Here we take a different approach.  We consider only a single scalar
field, but we use a weaker and, we hope, more physical condition on
the results of the black hole evaporation.  In the end, our results
still support the claim that late-time radiation cannot restore the
purity of the state of an evaporating black hole.

We will also make a more general investigation of our new definition
of a localized state.  In the very low energy regime we will find the
surprising result that this definition gives rise to much higher
entropy than the same energy could support in a rigid box.  In the
low-energy limit we conjecture that this difference grows without
bound.

\section{The vacuum-bounded state}

\subsection{Black hole evaporation}

We start by considering a black hole formed from a pure
quantum-mechanical state of incoming matter.  To avoid any possible
complications of quantum gravity theory, we will look at the state
produced after the black hole has completely evaporated
\cite{preskill:review}.  Gravity should play no significant role in
this state, since the energy density should be small
everywhere.\footnote{If instead there are Planck-scale concentrations
of energy, then we would have a remnant theory, a possibility we are
explicitly not considering here.}  We can describe the final state as
follows: at large distances from the position of the black hole (taken
as the origin) there is outgoing Hawking radiation, which we are
assuming to be thermal and without internal correlations.  Within some
distance $R$ of the origin, there is some state of ordinary quantum
fields that could have correlations with the radiation emitted
earlier.  The distance $R$ is the distance that such information might
have propagated since unknown physics came into play.  Let us assume
that Hawking's semi-classical calculation is good up to an energy
scale $\Tunk$.  This temperature is reached when the black hole's mass
is\footnote{We are working with units in which $c = G =
\hbar = k_B = 1$} $M_0 = 1/(8\pi \Tunk)$.\notationlabel{M_0} If, after
this, the rate of evaporation continues to match the Hawking
calculation,\footnote{As opposed, for example, to slowing to nothing
and leaving a remnant.}  the black hole will evaporate in time $t\sim
{10^4 M_0^3/g} \sim {1/(g
\Tunk^3)}$,
where \usenotation{g} is the
effective number of degrees of freedom in the particles that can be
radiated.  (See \cite{page:radiation}.)  So there is a sphere of radius
\be
R \sim {1\over g \Tunk^3}
\ee
which contains total energy
\be
\notationlabel{E_0}
E_0 = {1\over 8\pi \Tunk}
\ee
in which the information could be contained.
\label{sec:realistic-er}
Taking, for example, 
$\Tunk = 10^{15} \text{GeV} \sim 10^{-4} \mpl$,
and $g\sim 100$ we get\footnote{Another possibility is that $g$
diverges as $T\rightarrow \mpl$.  In this case the information can be
radiated in a small number of particles of about the Planck mass,
chosen from an infinite spectrum of such particles.  This is
effectively a remnant theory.}
\blea[goshwownumbers]
R & \sim & 10^{10}\lpl \sim 10^{-23}\text{cm}\\
E_0 & \sim & 10^3 \mpl \sim 10^{-2}\text{g} \sim 10^{19}\text{erg}\,.
\elea
This yields a somewhat outrageous density of $10^{65}\text{g}/\text{cm}^3$,
which is nevertheless small compared to the ``GUT
density'' $(10^{16} \text{GeV})^4/(\hbar^3 c^5) \sim
10^{81}\text{g}/\text{cm}^3$.

\subsection{The vacuum-bounded condition}

Now we would like to answer the following question: How much entropy
can be contained in a spherical region of radius $R$ with a total
energy $E_0$?  To answer this question we have to specify what we
mean by ``contained in a region.''  As mentioned earlier, if we ask
how much entropy can be contained in a spherical box of radius
$R$ with perfectly reflecting walls, the question can be easily
answered.  However, the system with the box is not so closely akin to
the system under discussion.  For instance, inserting the reflecting
walls into the system produces a divergent increase in the
ground-state energy of the system.  Furthermore, if we started with
the vacuum in the whole system, and then introduced a spherical wall,
we would produce a divergent geometric
entropy
\cite{bombelli:orig,srednicki:geom-ent,callan:geom-ent,holzhey:geom}.
A better description of our system is simply that it has thermal
radiation outside radius $R$, and an unknown state of the quantum
fields inside radius $R$, but no barrier or boundary at $R$.

To study such systems, we will assume that the difference between the
external Hawking radiation and an external vacuum is not important to
considerations of entropy.\footnote{If this approximation is bad we can
increase $R$ until the Hawking radiation outside $R$ has very
low temperature.}  We will study systems that have an arbitrary
state inside $R$ but the vacuum outside $R$.  To make this precise we
will specify the problem as follows:

\begin{quotation}
Let a {\em vacuum-bounded state} be a generalized state (i.e.\ density
matrix) for which every operator composed of fields at points outside
a specified interior region has the same expectation value as in the
vacuum.  What is the maximum entropy of such a state whose interior
region is a sphere of radius $R$\notationlabel{Rradius} and whose
total average energy\footnote{We cannot specify that every measurement
of the energy must give $E_0$.  Such a state is necessarily static and
thus cannot represent outgoing radiation.} is given by $\<H\> = E_0$?
\end{quotation}

We expect to find that the answer to this question is similar to that
of a box of radius $R$ with reflecting walls, with some small
correction.

We will denote by $X_\in$ quantities in the given interior region and
by $X_\out$ those outside this region.  We will say that a generalized
state is ``localized to the inside'' or ``obeys the vacuum-bounded
condition'' if any measurement performed on the outside field
operators in this state yields the same result as in the vacuum, i.e.\
if
\be[vac-cond]				
\Tr \rho O_\out = \Tr \rho^\vac O_\out = \<0|O_\out|0\>
\ee
for every operator $O_\out$ that is constructed from field operators in
the outside region.

In the language of density matrices, we can write $\rho_\out =
\Trin\rho$, where $\rho$ is the overall density matrix describing our
system and $\Trin$ means to trace over all the ``inside'' variables.
Then $\rho_\out$ is the reduced density matrix describing only the
outside variables, and \eqref{vac-cond} is equivalent to 
\be
\rho_\out = \rho_\out^\vac \equiv \Trin|0\>\<0|\,,
\ee
where $|0\>$ denotes the ground state.

\chapter{Mathematical Considerations}

In this chapter we will prove some theorems regarding the general
theory of constrained density operators.  Our goals are to show that
there is always a unique operator which maximizes the entropy subject
to the constraints and to demonstrate the form of that operator.
Unfortunately we will not accomplish either of these goals completely,
although we will make some progress in these directions.  In
particular we will show that if such a operator exists it is unique,
and that an operator of the form specified, if one exists, does in
fact maximize the entropy.

\section{Topology of the space of density operators}

\begin{definition}
Let $\calh$ be a separable Hilbert space, and let $\rho$ be a linear 
operator on $\calh$.  Then $\rho$\notationlabel{rho}
is a density operator (or density matrix) if and only if
\begin{itemize}\itemsep 0pt
\item $\rho$ is Hermitian
\item $\rho$ is trace-class and $\Tr\rho = 1$
\item $\<\psi|\rho|\psi\> \ge 0$ for any state $|\psi\>$.
\end{itemize}
\end{definition}
The last property is usually written ``$\rho$ is
positive semidefinite'' (or sometimes ``positive definite'') by
physicists and ``$\rho$ is positive'' or ``$\rho \ge
0$''\notationlabel{posdef}  by mathematicians.  In this section we
will usually use the latter notation.

We will require our operators $\rho$ to satisfy a set of constraints of the form
\be
\Tr\rho\, C_\alpha = V_\alpha
\notationlabel{Calpha}\notationlabel{Valpha}
\ee
for some given operators $C_\alpha$ and numbers $V_\alpha$.  (By $\Tr
A = a$ we mean that $A$ is trace-class and its trace is $a$.)  We
would like to find the density operator $\rho$ which maximizes the
entropy $S=-\Tr\rho\ln\rho$ subject to the constraints.  There are two
potential problems here.  First, there might be states with
arbitrarily large entropy.  If there are no constraints, that will be
the case whenever $\calh$ has infinite dimension.  Second, there might
be a supremum of the possible values of $S$ which is not achieved by
any $\rho$.

In an attempt to address these problems we would like to prove some
continuity results for $S$ and some compactness results for the space
of allowable $\rho$.  However, if any density operator is allowed, then
those with infinite entropy or infinite energy are dense in the space
of all density operators, using any reasonable norm.  Thus $S$ cannot
be continuous on such a space.  Instead we will work only with those
$\rho$ for which $\Tr\rho H$ is bounded, where $H$ is some reasonable
Hamiltonian.

Since density operators are trace-class they are also Hilbert-Schmidt
operators, and we can use the Hilbert-Schmidt inner product $(\rho,
\rho') = \Tr \rho\rho'$ and the corresponding norm given by
$\|\rho\|_2 = \sqrt{\Tr\rho^2}$.  The topology on the space of
operators will always be that induced by $\thenorm_2$.  We will write
$\thenorm$\notationlabel{norm}\notationlabel{opnorm} for the norm of a
state and for the bound on a bounded operator, and
$\thenorm_2$\notationlabel{hsnorm} for the Hilbert-Schmidt norm of an
operator.

\subsection{Compactness}

We will first show that with an energy bound the space of density
operators is compact.

\begin{theorem}\label{thm:compact}
Let $\calh$ be a separable Hilbert space, let $H$\notationlabel{Harb}
be a positive Hermitian operator on $\calh$ with a purely discrete
spectrum, and suppose further that no infinite-dimensional subspace
exists on which $H$ is bounded.\footnote{If such an
infinite-dimensional subspace does exist then there are states with
infinite entropy for any $E$.}  For some $E>0$\notationlabel{E} let
$\calp(E)$ be the space of density operators $\rho$ on $\calh$ such
that $\Tr\rho H\le E$.  Then the space $\calp$ is is compact in the
topology induced by $\thenorm_2$.
\end{theorem}

The proof is composed of a series of lemmas.

\begin{lemma}\label{lem:elts-converge}
Let $\{A_n\}$\notationlabel{A}\notationlabel{Asup} be a sequence of
operators on $\calh$ and suppose that $lim_{n\rightarrow\infty}
\Asup{n} = \bar A$.\notationlabel{Abar} Then 
$\lim_{n\rightarrow\infty} \Asup{n}_{\alpha\beta} = \bar A_{\alpha\beta}$
where $A_{\alpha\beta}$ denotes a matrix element of $A$ in some
orthonormal basis
$\{|\alpha\>\}$.
\end{lemma}

{\bf Proof:} If $\Asup{n} \rightarrow \bar A$ then $\| \Asup{n} - \bar
A \|_2 \rightarrow 0$.  We can use the basis $\{|\alpha\>\}$ to compute
the norm,
{
\setlength\abovedisplayskip{0pt}
\setlength\belowdisplayskip{0pt}
\be
| \Asup{n} - \bar A \|_2 =
\sum_{\alpha\beta} | \<\alpha|\Asup{n} - \bar A|\beta\> |^2 = 
\sum_{\alpha\beta} |\Asup{n}_{\alpha\beta} - \bar A_{\alpha\beta}|^2\,,
\ee}%
so each $|A_{\alpha\beta} - \Asup{n}_{\alpha\beta}| \rightarrow 0$
and thus $\Asup{n}_{\alpha\beta} \rightarrow \bar A_{\alpha\beta}$.\qed

\begin{lemma}\label{lem:pos-def}
Let $\{\Asup{n}\}$ be a sequence of positive operators on $\calh$ with
$\Asup{n}\rightarrow \bar A$.  Then $\bar A \ge 0$.
\end{lemma}

{\bf Proof:} Let $|x'\>$ be any element of $\calh$.  Define a
normalized vector $|x\> =
|x'\>/\| |x'\> \|$.  We can always make a basis with $|x\>$ as one of
its elements.  From \lemref{elts-converge}, $\Asup{n}_{xx}
\rightarrow \bar A_{xx}$. Thus
\be
\<x|\bar A|x\> = \lim_{n\rightarrow\infty} \<x|\Asup{n}|x\>\,.
\ee
Since $\Asup{n} \ge 0$ we have $\<x|\Asup{n}|x\> \ge 0$ for any
$|x\>$.  Since the limit of positive numbers cannot be negative,
$\<x|\bar A|x\> \ge 0$ and thus $\<x'|\bar A|x'\> \ge 0$
for any $|x'\>$.  That is to say, $\bar A \ge 0$.\qed

\begin{lemma}\label{lem:unbounded}
Let $B$ be any positive operator on $\calh$ with a purely discrete
spectrum.  Let $\{\Asup{n}\}$ be a sequence of positive operators with
$\Asup{n}\rightarrow \bar A$.  Suppose that $\Tr\Asup{n} B \le B_0$
for all $n$.  Then $\Tr\bar A B \le B_0$.
\end{lemma}

{\bf Proof:} We work in a basis where $B$ is diagonal.  We write the
trace
\be[Bsum-limit-1]
\Tr\bar A B = \lim_{N\rightarrow\infty}\sum_{n=1}^N\bar A_{nn} B_n\,.
\ee
For each $N$ this is a finite sum.  By \lemref{elts-converge}
the elements of $\Asup{i}$ converge individually to the elements of
$\bar A$.  Thus
\be[Bsum-limit-2]
\sum_{n=1}^N\bar A_{nn} B_n
= \lim_{i\rightarrow\infty} \sum_{n=1}^N\Asup{i}_{nn} B_n\,.
\ee
Since $\Asup{i} \ge 0$, the diagonal elements of $\Asup{i}$
cannot be negative, and since $B\ge 0$ we have $B_n\ge 0$ for every $n$.
  Thus
\be[Bsum]
\sum_{n=1}^N\Asup{i}_{nn} B_n \le \Tr \Asup{i} B \le B_0\,.
\ee
By \eqref{Bsum} the terms in the limit in \eqref{Bsum-limit-2} are
each no more than $B_0$, and thus
\be
\sum_{n=1}^N\bar A_{nn} B_n \le B_0\,.
\ee
Thus each term in the limit in \eqref{Bsum-limit-1} is not more than
$B_0$ and so $\Tr\bar A B \le B_0$.\qed

Note that \lemref{unbounded} would not hold with $\le$ replaced
by $=$, as shown by the following counterexample:  Let $B =
\diag(1,2,3,4\ldots)$.  For any $n$ let $\Asup{n} = 0$ except for the
$n$th diagonal element which is $1/n$.  By making $n$ large we can
make $\|\Asup{n}\|_2 = 1/n$ as small as desired.  Thus
$\Asup{n}\rightarrow\bar A = 0$.  However, $\Tr\Asup{n} B = 1$ for all
$n$, whereas $\Tr\bar A B = 0$.

We will now work in a basis $\{|n\>\}$ where $H$ is diagonal, 
$H_{mn} = \<m|H|n\> = \delta_{mn} E_n$ with $E_n \le E_{n+1}$.

\begin{lemma}\label{lem:smallsubtrace}
Let $A$ be any positive operator with $\Tr A H \le E$.  Given any
$\epsilon > 0$ there is some number $N$ such that for every positive
operator $A$ with $\Tr A H \le E$,
\be
\sum_{n=N}^\infty \<n|A|n\> \le \epsilon\,.
\ee
\end{lemma}

Note that $\Tr A H \le E$ implies that $A$ is trace-class and
consequently that for any given $A$, there is an $N$ with
$\sum_{N}^\infty A_{nn} \le \epsilon$.  The point of
\lemref{smallsubtrace} is that such an $N$ can be chosen uniformly for all
$A$.

{\bf Proof:}  We have
\be
\sum_{n=1}^\infty A_{nn} E_n = \Tr A H \le  E\,.
\ee
Since $ A \ge 0$ we have $ A_{nn} \ge 0$ for every $n$.  Since the $E_n$
are increasing,
\be
\sum_{n=N}^\infty  A_{nn} E_N < \sum_{n=N}^\infty A_{nn} E_n \le E
\ee
so
\be
\sum_{n=N}^\infty  A_{nn} < {E\over E_N}\,.
\ee
Since $H$ is unbounded on any infinite-dimensional subspace, we can
find $E_N$ arbitrarily large.  Thus for any $\epsilon>0$ we can find
an $N$ such that
\be[smallsubtrace]
\sum_{n=N}^\infty  A_{nn} < \epsilon
\ee
for every $A$ with $\Tr A H \le E$.\qed

\begin{lemma}\label{lem:norm-to-trace}
Given any $\epsilon > 0$ there is a $\delta > 0$ such that for any
positive operators $A$ and $A'$ with $\Tr A H \le E$ and $\Tr A H' \le
E$, if $\|A - A'\|_2 < \delta$ then $|\Tr(A-A')| < \epsilon$.
\end{lemma}

{\bf Proof:}
Let $\rho\downn N$ denote the $N\times N$ matrix given by
$\rho_{ab}$ with $a,b =1\ldots N$.
By \lemref{smallsubtrace} we can find $N$ such that
\be[norm-trace-1]
\Tr(A-A\downn N) = \sum_{n=N}^\infty A_{nn} < {\epsilon\over 3}
\ee
for any positive operator $A$ with $\Tr A H \le E$.  Let
\be
\delta = {\epsilon\over 3\sqrt{N}}\,.
\ee
Then for any $A$ and $A'$ with $\|A-A'\| < \delta$ we have
\be
\left(\Tr(A\downn N- A'\downn N)\right)^2
= \left(\sum_{n=1}^N (A_{nn} - A'_{nn})\right)^2
\le N \sum_{a=1}^N (A_{nn}-A'_{nn})^2
\!\le N \| A - A' \|_2^2 < {\epsilon^2\over 9}\,,
\ee
so
\be[norm-trace-2]
\left|\Tr(A\downn N-A'\downn N)\right| < {\epsilon\over 3}\,.
\ee
If $\Tr A H \le E$ and $\Tr A' H \le E$ we can add
\eqsref{norm-trace-1} and (\ref{eqn:norm-trace-2}) to get
\be
|\Tr A - \Tr A'| \le
|\Tr(A - A\downn N)| + |\Tr(A\downn N - A'\downn N)| + 
|\Tr(A'\downn N - A')| < \epsilon\,.\qed
\ee

\begin{lemma}\label{lem:trace-and-bounded}
Let $A$ be a trace-class operator and $C$ be a bounded operator.  Then
$AC$ is trace-class and $\Tr |A|C \le |\Tr A|\times\|C\|$.
\end{lemma}

{\bf Proof:} Since $A$ is trace-class it has a purely discrete spectrum.  Thus
we can write $A=\sum A_n|n\>\<n|$ with $\<n|n\> = 1$.  By the triangle
inequality, $\<n|C|n\> \le \||n\>\|\times \|C|n\>\| \le \|C\|$.  Then
\be
\Tr |A|C = \sum_n |A_n| \<n|C|n\> \le \sum_n \left(|A_n|\times\|C\|\right)
 = |\Tr A|\times\|C\|\,.\qed
\ee 

\begin{lemma}\label{lem:bounded}
Let $\{\Asup{i}\}$ be a sequence of positive trace-class operators with
$\Asup{i}\rightarrow \bar A$ and let $C$ be any bounded operator.
If $\Tr\Asup{i} C = c$ for all $i$ then $\Tr\bar A C = c$.
\end{lemma}

Compare with \lemref{unbounded}.  In the case of a bounded operator,
taking the limit preserves the trace exactly, whereas with an
unbounded operator we get only an upper bound.

{\bf Proof:} Since $\Asup{i}\rightarrow\bar A$ and using
\lemref{norm-to-trace}, for any $\epsilon > 0$ we can find $\|
\Asup{i} - \bar A \|_2$ sufficiently small that
\be
|\Tr (\Asup{i} - \bar A)| < {\epsilon\over \|C\|}\,.
\ee
Then by \lemref{trace-and-bounded}, $\Tr |\Asup{i} - \bar A| C <
\epsilon$.  Since $\Tr \Asup{i} C = c$, it follows that $|\Tr\bar A C
-c| < \epsilon$ for every $\epsilon > 0$ and thus that $\Tr\bar A C =
c$ as desired.

\begin{lemma}\label{lem:complete}
The space $\calp(E)$ is complete.
\end{lemma}

{\bf Proof:} Let $\{\rhosup{i}\}$ be a Cauchy sequence of density
operators in $\calp(E)$.  The space $\calp(E)$ is part of the space of
Hilbert-Schmidt operators on $\calh$, which is a complete space under
the $\thenorm_2$ norm.  Thus there is a Hilbert-Schmidt operator
\be
\rhobar = \lim_{i\rightarrow\infty}\rhosup{i}\,
\ee
Clearly $\rhobar$ is Hermitian.  By \lemref{pos-def}, $\rhobar \ge 0$.
By \lemref{unbounded}, $\Tr\rhobar H \le E$.  Then by
\lemref{bounded} with $C=I$, $\Tr\rhobar = 1$.  Thus $\rhobar \member
\calp(E)$.  So any Cauchy sequence in $\calp(E)$ converges to a limit in
$\calp(E)$, which is to say that $\calp(E)$ is complete.\qed

\begin{lemma}\label{lem:totally-bounded}
The space $\calp(E)$ is totally bounded, i.e.\ any sequence of $\rhosup{n}$
in $\calp(E)$ has a Cauchy subsequence.
\end{lemma}

{\bf Proof:} Let $\{\rhosup{n}\}$,
$n=1\ldots\infty$\notationlabel{rhosup} be an infinite sequence of
operators in $\calp(E)$.  As before we work in a basis where $H$ is
diagonal.  The space of Hermitian, positive, unit-trace $n\times n$
matrices is compact.  Thus any sequence of such matrices has a Cauchy
subsequence.  So define a set of sequences of integers $\{n_k(i)\}$,
$k\ge 0$ as follows: Let $n_0(i) = i$ and for each $k$ define
$\{n_k(i)\}$ to be a subsequence of $\{n_{k-1}(i)\}$ such that
$\rhosup{n_k(i)}\downn k$ is a Cauchy sequence.  Then let
$\hat\rhosup{i} = \rhosup{n_i(i)}$.\notationlabel{rhosuphat} We claim
that $\hat\rhosup{i}$ is a Cauchy sequence.

Given any $\epsilon>0$, by \lemref{smallsubtrace} we can find an $N$
such that
\be
\sum_{n=N}^\infty \rho_{nn} < {\epsilon\over 16}
\ee
for every $\rho \member \calp(E)$.
Since $\rho$ is positive,
$| \rho_{ab}|^2 <  \rho_{nn}  \rho_{bb}$.  Thus
\be
\sum_{a=1}^\infty\sum_{b=N}^\infty
|\rho_{ab}|^2 \le \sum_{a=1}^\infty\rho_{nn}\sum_{b=N}^\infty\rho_{bb}
\le \Tr\rho\cdot {\epsilon\over 16} = {\epsilon\over 16}\,.
\ee

Now since $\{\rhosup{n_N(i)}\downn N\}$ is a Cauchy sequence we can 
choose a number $M \ge N$ such that
\be
\|\rhosup{n_N(i)}\downn N - \rhosup{n_N(j)}\downn N\|_2^2 < \epsilon/2
\ee
for all $i,j \ge M$.  Then
\bea
\lefteqn{\left\|\rhosup{n_N(i)} - \rhosup{n_N(j)}\right\|_2^2
	- \left\|\rhosup{n_N(i)}\downn N - \rhosup{n_N(j)}\downn N\right\|_2^2} \nonumber\\
\hskip 10pt&=& \sum_{\hbox{$a$ or $b > N$}}
 \left|\rhosup{n_N(i)}_{ab} - \rhosup{n_N(j)}_{ab}\right|^2\nonumber\\
&\le & 2 \sum_{a=1}^\infty\sum_{b=N+1}^\infty 
   \left|\rhosup{n_N(i)}_{ab} - \rhosup{n_N(j)}_{ab}\right|^2\nonumber\\
&\le & 4 \sum_{a=1}^\infty\sum_{b=N+1}^\infty 
   \left(\left|\rhosup{n_N(i)}_{ab}\right|^2
    + \left|\rhosup{n_N(j)}_{ab}\right|^2\right)
< {\epsilon\over 2}\,.
\eea
Thus 
\be
\left\|\rhosup{n_N(i)} - \rhosup{n_N(j)}\right\|_2^2 \le \epsilon
\ee
for all $i,j \ge M$.  For any given $k \ge N$, $n_k(k)$ is $n_N(i)$
for some $i \ge k$, since the $n_k(i)$ are a subsequence of the
$n_N(i)$.  Thus for any $k, l \ge N$, and consequently for any $k, l
\ge M$, we find that
\be
\left\|\hat\rhosup{k} - \hat\rhosup{l}\right\|_2^2
= \left\|\rhosup{n_k(k)} - \rhosup{n_l(l)}\right\|_2^2
= \left\|\rhosup{n_N(i)} - \rhosup{n_N(j)}\right\|_2^2
\ee
for some $i, j \ge M$, and thus that
\be
\left\|\hat\rhosup{k} - \hat\rhosup{l}\right\|_2^2 < \epsilon
\ee
which is to say that $\hat\rhosup{i}$ is a Cauchy sequence.\qed

{\bf Proof of \thmref{compact}:} By \lemref{totally-bounded},
$\calp(E)$ is totally bounded.  By \lemref{complete}, $\calp(E)$
is complete, which is to say that $\calp(E)$ is compact.\qed

\subsection{Continuity of the entropy}

We now prove that $S$ is continuous if we restrict ourselves to $\rho$
with $\Tr\rho H \le E$.

\begin{theorem}\label{thm:scontinuous}
$S(\rho) = -\Tr\rho\ln\rho$ is continuous on $\calp(E)$ in the
$\thenorm_2$ norm.
\end{theorem}

{\bf Proof:} Let $\rho$ be any density operator in $\calp(E)$.  Given
any $\epsilon > 0$ we need to find a $\delta > 0$ such that $|S(\rho)
- S(\rho')| < \epsilon$ for every $\rho'\member\calp(E)$ with $\|\rho
- \rho'\|_2< \delta$.  We will do this by writing $\rho$ as the limit
of $N\times N$ matrices $\rho\downn N$ and $\rho'$ as the limit of
$\rho'\downn N$.

We work again in a basis where $H$ is diagonal.  Since $\rho\downn N$
is a matrix of finite dimension, $S(\rho\downn N)$ is continuous.
Thus given any $N$ we can find a $\delta$ such that
\be[S-finite-close]
|S(\rho\downn N) - S(\rho'\downn N)| < {\epsilon\over 3}
\ee
whenever $\|\rho\downn N - \rho'\downn N \|_2 < \delta$.
Thus it is sufficient to find an $N$ such that $|S(\rho\downn N) -
S(\rho)| < \epsilon/3$ for all $\rho \member \calp(E)$.

Let $p_k(\rho)$\notationlabel{pk} denote the $k$th largest eigenvalue
of $\rho$.  Lieb, Ruskai and Simon \cite{lieb:proof} show that
$p_k(\rho\downn N) \le p_k(\rho)$, and go on to show from this that
\be
\lim_{N\rightarrow\infty} S(\rho\downn N) = S(\rho)\,.
\ee
We want to extend this result to show that $S(\rho\downn N)
\rightarrow S(\rho)$ uniformly for all $\rho$ such that $\Tr\rho H \le E$.

Define the function
\be[sdefined]\notationlabel{s(p)}
s(p) = \cases{- p \ln p & $p>0$\cr
  		0 & $p=0$\,.\cr}
\ee
Note that $s$ is continuous. It is increasing for $p< 1/e$ and
decreasing for $p> 1/e$.  As we go from $S(\rho\downn N)$ to $S(\rho)$
the $p_k$ will increase.  It is possible to have a few large $p_k$
such that increasing $p_k$ will decrease $s(p_k)$, but for the
majority of $p_k$ an increase in $p_k$ means an increase in $s(p_k)$.

First we study how much $S$ can decrease as $N\rightarrow\infty$.  Let
$\epsilon$ be any small positive number.  By \lemref{smallsubtrace} we
can find $N_-$ large enough so that
\be
\sum_{a={N_-}}^\infty \rho_{nn} < {\epsilon\over 3}
\ee
for every $\rho\member\calp(E)$. Thus
\be
\Tr\rho\downn N = 1 - \sum_{n=N+1}^\infty \rho_{nn} > 1-{\epsilon\over 3}
\ee
for any $N > N_-$.
Now $\sum p_k(\rho) = \Tr\rho = 1$, so
\be
\sum_k \left(p_k(\rho) - p_k(\rho\downn N)\right) < {\epsilon\over 3}\,.
\ee
Since $ds/dp \ge -1$ for all $p\le 1$, we can conclude that
\be[S-decrease]
S(\rho\downn N) - S(\rho) < {\epsilon\over 3}
\ee
for all $N > N_-$.

Now we study how much $S$ can increase as $N\rightarrow\infty$.
We write
\be
\rho = \splitmatrix(\rho\downn N , A, A^T, \rho\upn N)\,.
\ee
For fixed $\rho\downn N$ and $\rho\upn N$, $S(\rho)$ will be largest
when $A=0$.  To see this, consider
\be
\rho_- = \splitmatrix(\rho\downn N, -A, -A^T, \rho\upn N)\,.
\ee
The eigenvalues of $\rho_-$ are the same as those of $\rho$, so
$S(\rho) = S(\rho_-)$.  Now let 
\be
\rho_0 = \splitmatrix(\rho\downn N, 0, 0, \rho\upn N)
= {\rho + \rho_-\over 2}\,.
\ee
Since the entropy is convex (see \cite{lieb:convex} and
\secref{convex}), $S(\rho_0) > (S(\rho) + S(\rho_-))/2 = S(\rho)$.
Thus $S(\rho) < S(\rho_0) = S(\rho\downn N) + S(\rho\upn N)$.

Now $\rho\upn N$ obeys the
constraints
\blea
\Tr \rho\upn N &=& 1-\Tr\rho\downn N \equiv 1- p_<\\
\Tr \rho\upn N H &\le&  E - \Tr \rho\downn N H\equiv E_>\,.
\elea
Let $\Sth(E ; e_1, e_2, e_3, \ldots)$ denote the thermal entropy of a system
whose energy levels are $e_1, e_2, e_3, \ldots.$
Suppose we had a system with energy levels $0$, $E_{N+1}$, $E_{N+2},
\ldots.$  
Then a possible normalized density matrix would be
\be
\splitmatrixul(p_<, 0, 0, \rho\upn N)\,.
\ee
It would have energy $E_> < E$ and entropy $- p_< \ln p_< +
S(\rho\upn N)$.
This must be less than the maximum entropy for this system with energy
$\le E$, so we conclude that
\be
S(\rho) - S(\rho\downn N) < S(\rho\upn N)
< \Sth(E ; 0, E_{N+1}, E_{N+2}, \ldots)\,.
\ee
Now as $N\rightarrow\infty$, only higher and higher energy states are
permitted in the expression above, so the entropy will decrease to
zero.  That is,
\be
\lim_{N\rightarrow\infty} \Sth(E ; 0, E_{N+1}, E_{N+2}, \ldots) = 0\,.
\ee
So, given $\epsilon > 0$ we can find $N_+$ such that if $N > N_+$
then
\be[S-increase]
S(\rho) - S(\rho\downn N) < {\epsilon\over 3}
\ee
for any $\rho\member\calp(E)$.  

Now we set $N = \max(N_+, N_-)$.   From \eqsref{S-decrease} and
(\ref{eqn:S-increase}) we find that
\be
|S(\rho) - S(\rho\downn N)| < {\epsilon\over 3}\,,
\ee
for every $\rho\member\calp(E)$.  Then we choose $\delta$ as in
\eqref{S-finite-close}. If $\|\rho - \rho'\|_2 < \delta$ then 
\be
|S(\rho)- S(\rho')| < |S(\rho) - S(\rho\downn N)|
+ |S(\rho\downn N) - S(\rho'\downn N)|
+ |S(\rho'\downn N) - S(\rho')|
< \epsilon\,.\qed
\ee

\begin{theorem}\label{thm:manypositive}
Let $\calh$ and $H$ be defined as in \thmref{compact}, let
$\{C_\alpha\}$ be a set of positive operators with discrete spectra,
and let $\calpmanyle$ be the set of density operators $\rho$ such that
$\Tr\rho H \le E$ and $\Tr\rho C_\alpha \le V_\alpha$ for all
$\alpha$.  Then $S(\rho)$ achieves a maximum on $\calpmanyle$.
\end{theorem}

{\bf Proof:} From \thmref{compact} we know that $\calp(E)$ is compact.
Thus any sequence of $\rhosup{n} \member \calpmanyle$ has a subsequence
that converges to some $\rhobar \member \calp(E)$.  By repeated
application of \lemref{unbounded} we see that $\Tr \rhobar
C_\alpha \le V_\alpha$ so $\rhobar \member \calpmanyle$.  Thus
$\calpmanyle$ is sequentially compact, and so it is compact.  By
\thmref{scontinuous}, $S$ is continuous on $\calp(E)$, so $S$ achieves
a maximum on $\calpmanyle$.\qed

It is possible to rearrange the constraints that we will need into
positive operators.  However, our constraints are of the type $\Tr\rho
C_\alpha =  V_\alpha$ rather than $\Tr\rho
C_\alpha \le V_\alpha$.  What we really need is

\begin{conjecture}\label{conj:goodanswer}
Define $\calh$, $H$ and $\{C_\alpha\}$ as in
theorem \ref{thm:manypositive} and let $\calpmanyeq$ be the set of density
operators $\rho$ such that $\Tr\rho H = E$ and $\Tr\rho C_\alpha =
V_\alpha$ for all $\alpha$.  Then $S(\rho)$ achieves a maximum
on $\calpmanyeq$.
\end{conjecture}
Unfortunately, we do not know how to complete the proof of this
conjecture.  However, in the next section we show that if there is
such a $\rho$ it is unique.  Furthermore since $\calpmanyeq \subset
\calpmanyle$ the maximum value of $S$ in \thmref{manypositive} is at
worst an upper bound on $S(\rho)$ with $\rho \member \calpmanyeq$.

\section{Maximization of the entropy}

To maximize the entropy we would like to look at the variation of $S$ as
$\rho$ is varied and require that $\ddelta S = 0$ so that $S$ can be a
maximum.  However, when $\rho$ is varied we may find that $\ddelta S$
is not well-defined.  There are two possible causes for this.  One is
that $\rho$ has some zero eigenvalues.  The second is that $\rho$ has
arbitrarily small eigenvalues which become negative under any
variation.  We treat these problems below.

First, since we are concerned with zero eigenvalues we should specify
what we mean by $S$.  When we write $S(\rho)=-\Tr\rho\ln\rho$ we mean
that $S(\rho)=s(\rho)$, where $s$ is defined in \eqref{sdefined}.
That is, if $\rho = \sum P_\alpha |\alpha\>\<\alpha|$ then $S = \sum
s(P_\alpha)$ or alternatively if we write $s(p)$ as
a power series in $p$ then $S(\rho)$ is given by the same power
series in $\rho$.  Under this definition $S(\rho)$ is well-defined
when $\rho$ has zero eigenvalues, even though $\ln\rho$ is not defined
in such cases.

\subsection{Variation of $S$}

We want to vary $\rho$ via $\rho' = \rho + t\ddelta\rho$ and look at
the resulting change in $S(\rho')$.  We will call a variation
$\ddelta\rho$ admissible if:
\begin{itemize}\itemsep 0pt
\item $\ddelta\rho$ is Hermitian.
\item $\Tr\ddelta\rho = 0$, so that $\Tr\rho' = 1$.
\item For each $\alpha$, $\Tr\ddelta\rho C_\alpha = 0$, so that $\Tr
\rho'C_\alpha$ is unchanged.
\item For sufficiently small $t>0$, $\rho' \ge 0$.
\end{itemize}
Under these conditions $\rho'$ is a well-defined density operator for at
least some small range of $t>0$.

We will often want to let $\ddelta\rho$ interpolate between $\rho$ and
some other density operator $\rhobar$.  To show that this is always
possible, we prove
\begin{theorem}\label{thm:rhospaceconvex}
The space of density operators $\rho$ that obey the constraints is
convex, i.e.\ if $\rho$ and $\rhobar$ are density operators that obey
the constraints then $\ddelta\rho = \rhobar - \rho$ is an admissible
variation and $\rho_0 = \rho + t\ddelta\rho$ is a density operator for
all $t\member [0,1]$.
\end{theorem}

{\bf Proof:} 
\begin{itemize}\itemsep 0pt
\item Since $\rho$ and $\rhobar$ are Hermitian, so is $\ddelta\rho$.
\item $\Tr\ddelta\rho = \Tr\rhobar - \Tr\rho = 0$.
\item $\Tr\ddelta\rho C_\alpha = \Tr\rhobar C_\alpha - \Tr\rho C_\alpha
	= 0$.
\item Let $\bx$ be any state vector.  Then for all $t\member
        [0,1]$, $\bx\cdot\rho'\bx 
	= \bx\cdot(t\rhobar+(1-t)\rho)\bx = t(\bx\cdot\rhobar\bx)
	+ (1-t)(\bx\cdot\rho\bx) \ge 0$, i.e.\ $\rho' \ge 0$.\qed
\end{itemize}

Now
\be
S(\rho) = \sum_\alpha s(p_\alpha)
\ee
where $p_\alpha$ are the eigenvalues of $\rho$.  When we vary $\rho$,
the change in the eigenvalues is given by first-order perturbation
theory,
\be
{dp_\alpha\over dt} = \<\alpha|\ddelta\rho|\alpha\> =
\ddelta\rho_{\alpha\alpha}\,.
\ee
If $p_\alpha > 0$ then we have
\be
{ds(p_\alpha)\over dp_\alpha} = -(1+\ln p_\alpha)\,.
\ee
Thus if there are no zero eigenvalues we can write
\be[dsdt]
{dS\over dt} = - \sum_\alpha (1+\ln p_\alpha) \ddelta\rho_\alpha
= -\Tr\ddelta\rho - \Tr\ddelta\rho\ln\rho'
= -\Tr\ddelta\rho\ln\rho'\,,
\ee
since we have required $\Tr\ddelta\rho = 0$.

\subsection{Prohibited sectors}\label{sec:prohib}

If there are zero eigenvalues, then $\ln\rho$ and thus \eqref{dsdt}
are not well-defined.  We handle this case as follows:

\begin{theorem}
If $\rho$ maximizes $S$ subject to the constraints $\Tr \rho C_\alpha =
V_\alpha$ and if $\rho|\alpha\> = 0$, then every $\rho$ that satisfies the
constraints must annihilate $|\alpha\>$.
\end{theorem}

{\bf Proof:} Suppose to the contrary that $\rho|\alpha\> = 0$ but
there is some $\bar\rho$ such that $\Tr \bar\rho C_\alpha = V_\alpha$ but
$\bar\rho|\alpha\> \neq 0$.  We let $\ddelta\rho = \bar\rho - \rho$.  By
theorem \ref{thm:rhospaceconvex}, $\ddelta\rho$ is admissible.
Now we look at just those $p_\alpha$ that are in fact changed by
$\ddelta\rho$.  Let $t$ be some small but positive number.  Then
all the $p_\alpha$ that are changed by $\ddelta\rho$ are positive and
we can write
\be[sumfords]
{dS\over dt} = 
	- \sum_{\hbox{$\alpha$ with $\ddelta\rho_\alpha \neq 0$}}
	\ddelta\rho_\alpha \ln\rho_\alpha\,.
\ee
As $t\rightarrow 0$ the $- \ln p_\alpha$ terms grow without bound for
those $\alpha$ where $p_\alpha(t) \rightarrow 0$.  Thus eventually these terms
dominate everything else in \eqref{sumfords} so that
\be
\lim_{t\rightarrow 0} {dS\over dt} = +\infty\,.
\ee
Thus for sufficiently small $t>0$ there is a $\bar\rho$ with $S(\bar\rho) >
S(\rho)$ in contradiction to assumption.  Thus if $\rho$ maximizes $S$
then we must have $\bar\rho|\alpha\>=0$ for every $|\alpha\>$ where
$\rho|\alpha\> = 0$.\qed

Thus $\rho$ can have a nontrivial nullspace $\caln\subset\calh$ only
if every every $\rho$ that satisfies the constraints annihilates the
space $\caln$.  In this case, we will write $\calh = \caln \oplus
\calh'$\notationlabel{calh'} and work in the space $\calh'$.  We will
replace all our operators with ones that act only on $\calh'$.  If,
restricted to $\calh'$, the constraints are linearly dependent, we can
now discard some of them to have again a linearly independent set.
This process eliminates the troublesome sector from the problem.

\subsection{Derivatives of $S$}

If $\ddelta\rho$ is an admissible variation, and as
long as there are no zero eigenvalues, we can differentiate $S$ with
respect to $t$ as in \eqref{dsdt}.  We demand that
\be[dsdtagain]
{dS\over dt} = -\Tr\ddelta\rho\ln\rho' = 0
\ee
for $S$ to be a maximum.

\label{sec:convex}
For the second derivative we care only about the sign.  Lieb
\cite{lieb:convex} showed that $S$ is always (downward) concave.  Here
we obtain the same result by a different technique and show also that
the concavity is strict, i.e.\ that 
\be[dsdt2neg]
{d^2S\over dt^2} < 0
\ee
for any $\rho$ and any nonzero $\ddelta\rho$.

From \eqref{dsdtagain}
\be
{d^2S \over dt^2} = - \Tr \ddelta\rho {d\over dt}\ln\rho'\,.
\ee
To expand this we use the formula
\be[dln]
{d\over dt}\ln A = \int_0^1 ds \left(I-s(A-I)\right)^{-1} {dA\over dt}
\left(I-s(A-I)\right)^{-1}
\ee
which can easily be derived as follows: Let $B = A - I$.  We expand
\be
\ln A = \ln(I+B) = B + {B^2\over 2} + {B^3\over 3} + \cdots
\ee
and differentiate to get
\bea
\lefteqn{{d\over dt} \ln A = {dB\over dt} +
{1\over 2}\left({dB\over dt}B + B{dB\over dt}\right)}\hspace{20pt}&\nonumber\\
&& + {1\over 3}\left({dB\over dt}B^2 + B{dB\over dt}B + B^2 {dB\over dt}\right)
+ \cdots\,.
\eea
We observe that this is
\bea
\lefteqn{\int_0^1 ds {dB\over dt} +
s\left({dB\over dt}B + B{dB\over dt}\right)
 + s^2\left({dB\over dt}B^2 + B{dB\over dt}B +
B^2 {dB\over dt}\right) }\hspace{10pt} & \nonumber\\
& = & \int_0^1 ds \left(1+sB+s^2B^2+\cdots\right) {dB\over dt}
     \left(1+sB+s^2B^2+\cdots\right) \nonumber\\
& = & \int_0^1 ds \left(I-sB\right)^{-1} {dB\over dt} 
         \left(I-sB\right)^{-1}
\eea
as desired.

Using \eqref{dln} we find
\bea
{d^2S\over dt^2} &=& - \int_0^1 ds \Tr \ddelta\rho (I-s(\rho-I))^{-1}
   \ddelta\rho \left(I-s(\rho-I)\right)^{-1}\nonumber\\
&=& - \int_0^1 \Tr X(s)^2
\eea
where
\be
X(s) \equiv (\ddelta\rho)^{1/2} (I-s(\rho-I))^{-1} (\ddelta\rho)^{1/2}\,.
\ee
Since $X(s)$ is Hermitian, $\Tr X(s)^2 \ge 0$ with equality obtained
only for $X(s) = 0$.  Thus
\be
{dS^2\over dt^2} \le 0
\ee
with equality only if $X(0) = 0$ for all $s$.  But $X(0) =
\ddelta\rho$, and thus
\be
{dS^2\over dt^2} < 0
\ee
for any nonzero $\ddelta\rho$.

Now suppose that $\rho$ maximizes $S$.  Then for any admissible
$\ddelta\rho$, we must have
\be[dsneg]
{dS\over dt} \le 0
\ee
or else for sufficiently small $t$ we would have $S(\rho') >
S(\rho)$.  Furthermore we can see that condition (\ref{eqn:dsneg}) is
sufficient to show that $\rho$ maximizes $S$, as follows:  Let
$\rhobar$ be any other density operator that meets the constraints.  Let
$\ddelta\rho = \rhobar - \rho$.  From theorem
\ref{thm:rhospaceconvex}, $\ddelta\rho$ is admissible.
From \eqsref{dsneg} and (\ref{eqn:dsdt2neg}), it follows that
$S(\rhobar) < S(\rho)$.  Since this holds for any $\rhobar$, $\rho$
is the global maximum of $S$.  It follows from this that if there is a
$\rho$ in conjecture \ref{conj:goodanswer} it is unique.
\label{sec:unique}

\section{The form of $\rho$}

We must find the unique state $\rho$ (if any) which satisfies our
constraints and which gives $\Tr\ddelta\rho\ln\rho = 0$ for any
$\ddelta\rho$ which maintains the constraints.
This means that we are concerned with $\ddelta\rho$ such that
\blea[drho]
\Tr\ddelta\rho & = & 0 \\
\Tr\ddelta\rho C_\alpha & = & 0
\elea
for all $\alpha$.  We have included the Hamiltonian among the
constraints $C_\alpha$.  For every $\ddelta\rho$ satisfying Eqs.\
(\ref{eqn:drho}) we must have
\be[lnrhocond]
\Tr\ddelta\rho\ln\rho = 0\,.
\ee
If we choose
\be
\ln\rho = {\text{const}} + \sum_\alpha f_\alpha C_\alpha
\ee
which is to say
\be[rhoform]
\rho \propto e^{\sum f_\alpha C_\alpha}\,,
\ee
with any coefficients $f_\alpha$, then \eqsref{drho} ensure
that \eqref{lnrhocond} is satisfied.  If we can find a $\rho$
with the form of \eqref{rhoform} that satisfies the constraints,
then we have found the unique solution.  We have not shown that
such a solution always exists, but in the numerical work
described in \secref{vb-computation} we have always succeeded in
finding one.

Since $\Tr\rho = 1$, we
can write
\be
\rho = {e^{\sum f_\alpha C_\alpha} \over \Tr e^{\sum f_\alpha C_\alpha}}\,.
\ee
Our goal is now to determine the coefficients $f_\alpha$ so that the
constraints are satisfied.

We can define a grand partition function,
\be
Q=\Tr e^{\sum f_\alpha C_\alpha}\,.
\notationlabel{Qgrand}
\ee
Its derivatives are
\be
{dQ\over df_\alpha} = \Tr C_\alpha e^{\sum f_\alpha C_\alpha}\,,
\ee
so
\be
\<C_\alpha\> = {d\over df_\alpha}\ln Q\,.
\ee
We have the usual thermodynamic formula for the entropy,
\be
S = -\<\ln\rho\> = \ln Q - \sum f_\alpha \<C_\alpha\>\,.
\ee
Differentiating this we find
\be
{dS\over df_\alpha} = - \sum f_\beta {d\<C_\beta\>\over df_\alpha}\,.
\ee

Now we specialize to the case where one of the constraints is
just the Hamiltonian.  The corresponding coefficient is written
$-\beta$\notationlabel{beta}, and we have
\be
\rho = {1\over Q}e^{-\beta H + \sum f_\alpha C_\alpha}\,.
\ee
If we vary the coefficients in such a way that $\<H\> = E$ changes but
the other $\<C_\alpha\> = V_\alpha$ remain fixed we see that
\be
{dS = - \sum f_\alpha d\<C_\alpha\> = \beta dE}\,.
\ee
Thus $\beta = dS/dE$ and so the coefficient $\beta$ has the usual
interpretation as the inverse temperature.

\chapter{The Problem}

Now we return to the problem at hand.  We would like to find the
maximum-entropy vacuum-bounded state with a given average energy
$E_0$.  We first reduce the problem to a simpler case which we hope
will capture the important behavior.  Then we analyze the sectors on
which $\rho$ must be zero and introduce new coordinates in which these
sectors do not appear.  We work with these new coordinates to derive
formulas for the expectation values of the operators that must agree
with the vacuum.  Finally we look at the form of the results to be
expected when we solve the problem numerically in the next chapter.

\section{Simplifications}

\subsection{One scalar field, one dimension}

First we restrict ourselves to a theory consisting only of gravity
and one massless scalar field.  In such a system we can imagine
preparing an incoming shell of scalar particles to form a black hole,
which would then evaporate by emitting scalar quanta.  Thus we can
ask the same questions about black hole evaporation in this system as
we could, say, in the standard model plus gravity.

We will also begin here by working in one dimension.  We will put our
entire system in a box of length $L$ and require that all deviations
from the vacuum are in the region from $0$ to $\Lin$.  Later we will
take the overall box size $L$ to infinity while $\Lin$ remains
fixed.  The inside region will be $[0, \Lin]$ and the outside region
will be $[\Lin, L]$.  We will use the usual scalar field Hamiltonian,
which in classical form is
\be
H = \half \int_0^L \left[\pi(x)^2 
 + \left({d\phi\over dx}\right)^2\right] dx\,.
\notationlabel{H}
\notationlabel{phi(x)}\notationlabel{pi(x)}
\ee

\subsection{Gaussian form}

In our problem, the constraints are the energy bound,
\be
\Tr\rho H = E_0\,,
\ee
and the vacuum-bounded condition,
\be
\Tr\rho O_\out^\alpha = \<0|O_\out^\alpha|0\>\,,
\ee
where $O_\out^\alpha$\notationlabel{Oout}
is any operator which is constructed out of the
fields $\phi(x)$ and $\pi(x)$ in the outside region.  We will assume
the solution has the form
\be
\rho \propto e^{-\beta H + \sum f_\alpha O_\out^\alpha}\,.
\ee
We now show that $f_\alpha$ is nonzero only for those
operators $O_\out^\alpha$ which are quadratic in the fields.

Suppose that we wanted to solve a different problem in which we cared
only about the constraints involving the quadratic operators.  We would
have the energy bound and the constraints
\be
\Tr\rho Q_\out^a = \<0|Q_\out^a|0\>\,,
\notationlabel{Qout}
\ee
where $Q_\out^a$ runs only over quadratic operators $\phi(x)\phi(y)$
and $\pi(x)\pi(y)$.  (The operators $\phi(x)\pi(y)$ vanish
automatically by symmetry under $\phi \rightarrow -\phi$.)  We expect
the solution to this problem to have the form
\be
\rho' \propto e^{-\beta H + \sum f_a Q_\out^a}\,.
\ee

Now $\rho'$ is a Gaussian operator; i.e., $\<\phi(\cdot)|\rho'|\phi'(\cdot)\>$
is a Gaussian functional of the values of $\phi$ and $\phi'$.  Let
$\rho'_\out = \Trin \rho'$.  The trace is just a set of Gaussian
integrals, which means that the resulting $\rho'_\out$ is also a
Gaussian.  Because $H$ is quadratic, the vacuum $\rho^\vac = |0\>\<0|$ is
Gaussian, and so is its
trace $\rho^\vac_\out = \Trin \rho^\vac$.  Now by construction we have
$\Tr\rho'_\out \phi(x)\phi(y) = \Tr\rho^\vac_\out \phi(x)\phi(y)$ and
$\Tr\rho'_\out \pi(x)\pi(y) = \Tr\rho^\vac_\out \pi(x)\pi(y)$.  These
conditions are sufficient to fix the coefficients in the Gaussian
$\rho'_\out$, and thus to show that in fact $\rho'_\out$ and $\rho^\vac_\out$ are the
same Gaussian; i.e.\ that $\Trin\rho' = \Trin\rho^\vac$.

Thus $\rho'$ satisfies all the constraints of the original problem.
Since only one $\rho$ can have these properties it follows that
$\rho = \rho'$ and thus that Gaussian solution $\rho'$ is the correct
solution to the original problem.

\subsection{The discrete case}

We now approximate the continuum by a one-dim\-en\-sional lattice of coupled
oscillators, with a classical Hamiltonian
\be[hdisc]
H = \half\left(\Px \cdot \Px + \bx \cdot K \bx\right)\,.
\notationlabel{bx}
\notationlabel{K}
\ee
The simple kinetic term in \eqref{hdisc} corresponds to
choosing oscillators of unit mass, regardless of how densely they are
packed.  In terms of scalar-field variables this means that $x_\mu =
\sqrt{L_1}\phi_\mu$ and $P_\mu = \pi_\mu/\sqrt{L_1}$ where $L_1 =
L/(N+1)$\notationlabel{L_1} is the lattice spacing, $\phi_\mu$ is the
average of $\phi(x)$ over an interval of length $L_1$, and $\pi_\mu$ is
the total momentum $\pi(x)$ in the interval.

The matrix $K$ gives the couplings between the oscillators and
represents the $d\phi/dx$ term in the scalar field Hamiltonian.  To
approximate the continuum with the zero-field boundary condition we
will imagine that we have $N$ oscillators located at the points
$1/(N+1)\ldots N/(N+1)$ and that the end oscillators are coupled to
fixed-zero oscillators at $0$ and $1$.  Then

\be
K = \left(\matrix{2g & -g & 0 & 0 & \hspace{5pt}\cdot\hspace{5pt} \cr
                    -g & 2g & -g & 0 & \cdot \cr
                    0 & -g & \cdot & \cdot & \cdot \cr
                    0 & 0 & \cdot & \cdot & \cdot \cr
                    \cdot & \cdot & \cdot & \cdot & \cdot \cr
                    }\right)
\ee
where $g = 1/L_1^2$.

We will take $\Nin \approx \Lin/L_1$ of the oscillators to represent
the inside region, and $\Nout = N - \Nin$ to represent the outside
region.

We want to maximize $S$ subject to the constraints
\blea[rhoconstraints]
\Tr\rho H & = & E_0\label{eqn:Hrhoconstraint}\\
\Tr\rho x_i x_j & = & \<0|x_i x_j|0\>\label{eqn:xrhoconstraint}\\ 
\Tr\rho P_i P_j & = & \<0|P_i P_j|0\>\label{eqn:Prhoconstraint}\,,
\elea
where $i$ and $j$ run over the oscillators which represent 
the outside region.
We will define matrices $\xop$ and $\pop$ whose elements are the
quadratic operators via
\blea
\xopsub_{\mu\nu} &=& x_\mu x_\nu\\
\popsub_{\mu\nu} &=& P_\mu P_\nu
\elea
so that \eqsref{xrhoconstraint} and (\ref{eqn:Prhoconstraint})
become
\blea
\Tr\rho\xop_\outout &=& \<0|\xop_\outout|0\>\\
\Tr\rho\pop_\outout &=& \<0|\pop_\outout|0\>\,.
\elea

\section{Prohibited sectors}

As discussed in \secref{prohib}, the first thing we need to
do is to look for sectors of our Hilbert space that any $\rho$ must
annihilate in order to meet the constraints, and restrict our
attention to the subspace orthogonal to these.  We will start by
examining the ground state of our system and looking for sectors which are
forced to remain in the ground state.

\subsection{A different description of the vacuum}

\label{sec:oldgroundstate}

To work with the Hamiltonian of \eqref{hdisc} we will make a change of
coordinate to put it in diagonal form.  Let \usenotation{Z} be a
matrix whose columns are the eigenvectors of $K$, $K = Z \Omega_0^2
Z^{-1}$, with the normalization
\be[Znorm]
Z\Omega_0 Z^T = I \qquad\hbox{and}\qquad Z^\invt \Omega_0 Z^{-1} = K\,.
\ee
Define new coordinates $\bz$ via
$\bx = Z \bz$ and $\Px = Z^\invt \Pz$.  In these coordinates,
\be
\notationlabel{omega0}
H = \half \sum_\alpha \omega^{(0)}_\alpha 
\left(P_{z_\alpha}^2 + z_\alpha^2\right)\,.
\ee
The vacuum is the ground state of this Hamiltonian.  We can define
raising and lowering operators
\blea
a_\alpha & = & {1\over\sqrt{2}}\left(z_\alpha+iP_{z_\alpha}\right)\\
\adag_\alpha & = & {1\over\sqrt{2}}\left(z_\alpha-iP_{z_\alpha}\right)\\
H & = & \sum_\alpha \omega^{(0)}_\alpha \left(\adag a + \half\right)\,.
\elea
The vacuum is the state $|0\>$ annihilated by all the $a_\alpha$.
It is straightforward to write the expectation values in the vacuum
state,
\blea
\zopexsub_{\alpha\beta} \equiv \<0|z_\alpha z_\beta|0\>= \half\delta_{\alpha\beta}\\
\pzopexsub_{\alpha\beta} \equiv \<0|P_{z_\alpha} P_{z_\beta}|0\>
= \half\delta_{\alpha\beta}
\elea
so
\blea[xxppdef]
\<0|\xop|0\> & = & Z \<0|\zop|0\> Z^T = \half Z Z^T\\
\<0|\pop|0\> & = & Z^\invt \<0|\pzop|0\> Z^{-1}= \half Z^\invt Z^{-1}
= {1\over 4}(\<0|\xop|0\>)^{-1}\label{eqn:xoppinv}\,.
\elea

There can be many different
Hamiltonians that have the same ground state.  If we consider
\be
\notationlabel{altH}
H' = \half\left(\Px \cdot T' \Px + \bx \cdot K' \bx\right)
\ee
with $T'$ and $K'$ some coupling
matrices, we can follow the
above derivation to get a normal mode matrix \usenotation{Y} and some
frequencies $\Omega$\notationlabel{Omega} with
\blea
\notationlabel{omega}
Y \Omega Y^T & = & T'\\
Y^\invt \Omega Y^{-1} & = & K'\\
\bx & = & Y \by\\
\Px & = & Y^\invt \Py\\
H' & = & \half \sum_\beta \omega_\beta 
\left(P_{y_\beta}^2 + y_\beta^2\right)\,.
\elea
We can define raising and lowering operators for these modes,
\blea
b_\beta & = & {1\over\sqrt{2}}\left(y_\beta+iP_{y_\beta}\right)\\
\bdag_\beta & = & {1\over\sqrt{2}}\left(y_\beta-iP_{y_\beta}\right)\\
H' & = & \sum_\beta \omega_\beta \left(\bdag b + \half\right)\,.
\elea

Now $\by = Y^{-1} \bx = Y^{-1} Z \bz = W^{-1} \bz$ where
\be
W \equiv Z^{-1} Y\,.
\notationlabel{W}
\ee
Similarly, $\Py = Y^T\Px = Y^T Z^\invt \Pz = W^T \Pz$.
Consequently,
\bea
b_\beta & = & {1\over\sqrt{2}}\left(W^{-1}_{\beta\alpha}z_\alpha
+ iW^T_{\beta\alpha}P_{z_\alpha}\right)\nonumber\\
& = & \half\left(W^{-1}_{\beta\alpha}(a_\alpha + \adag_\alpha)
+ W^T_{\beta\alpha}(a_\alpha - \adag_\alpha)\right)\nonumber\\
& = & \half\left((W^{-1}+W^T)_{\beta\alpha}a_\alpha
+ (W^{-1} - W^T)_{\beta\alpha}\adag_\alpha\right) \,.
\eea
For $H$ and $H'$ to have the same vacuum we require that the $b_\beta$
depend only on the $a_\alpha$ and not on the $\adag_\alpha$, which is
to say that $W^{-1} = W^T$, i.e.\ that $W$ is a unitary matrix. With
$W$ unitary, $Y Y^T = Z Z^T$ so the vacuum expectation values of
\eqref{xxppdef} have the same values expressed in terms of $Y$ as they
had in terms of $Z$.

\subsection{Modes that remain in the ground state}

Now consider a unitary matrix $W$ and let $Y = Z W$ as in the last
section.  Suppose we can find $W$ such that $Y$ has the following
property:
\begin{quote}
The $N$ modes can be divided into $\Ngs > 0$ ``ground state'' modes
and $\Nfree \equiv N - \Ngs$ ``free'' modes such that for all $a \le
\Nin$ and for all $\beta > \Nfree$, $Y_{a\beta} = 0$ and
$Y^{-1}_{\beta a} = 0$.
\end{quote}
That is to say $Y$ and $Y^{-1}$ will have the form
\be[Yform]
Y = \sizedsplitmatrix(Y_\infree, 0, Y_\outfree, Y_\outgs)(\Nfree, \Ngs)(\Nin, \Nout)
\ee
and
\be[Yinvform]
Y^{-1} = \sizedsplitmatrix(Y^{-1}_\freein, Y^{-1}_\freeout, 0,
			Y^{-1}_\gsout)(\Nin, \Nout)(\Nfree,\Ngs)
\,.
\ee

If this is the case, then for $\beta > \Nfree$,
$b_\beta = \left(Y^{-1}_{\beta\gamma}x_\gamma
+ i Y_{\gamma\beta} P_{x_\gamma}\right)/\sqrt{2}$ depends only on outside
operators $x_i$ and $P_{x_i}$.  In any vacuum-bounded state,
regardless of entropy considerations, $\<x_i x_j\>$ and $\<P_iP_j\>$
have their vacuum values.  Consequently, for all $\beta > \Nfree$, if
$\rho$ describes a vacuum-bounded state then $\Tr\rho\bdag_\beta
b_\beta = \<0|\bdag_\beta b_\beta|0\> = 0$.  The vacuum-bounded
constraint forces modes $\beta > \Nfree$ to be in their ground states.
These modes will not contribute to the calculation of the
maximum-entropy vacuum-bounded state.

How many such modes can exist?  Let $W_\beta$ denote a column of $W$
with $\beta > \Nfree$ and let $Z^a$ denote a row of $Z$ with $a
\le \Nin$.  Similarly let $(Z^{-1})_a$ denote a column of
$Z^{-1}$.  Since $Y = Z W$, $Y_{a \beta} = 0$ whenever $W_\beta$
is orthogonal to $Z^a$.  Similarly $Y^{-1} = W^T Z^{-1}$ so
$Y^{-1}_{\beta a} = 0$ whenever $W_\beta$ is orthogonal to
$(Z^{-1})_a$.  Since $W$ is unitary, the $W_\beta$ must also be
orthogonal to each other.  Thus there are $\Ngs$ columns of $W$ which have
to be orthogonal to $\Nin$ rows of
$Z$, to $\Nin$ columns of $Z^{-1}$, and to each other.  Since
there are $N$ components in a column of $W$ it can be orthogonal in
general to at most $N-1$ other vectors.  Thus $\Ngs$ is limited by
$N-1 = 2\Nin + \Ngs -1$, or $\Ngs = N - 2\Nin = \Nout - \Nin$.  Thus
whenever $\Nout > \Nin$ there will be $\Ngs = \Nout - \Nin$ modes that are
forced to remain in the ground state.

These conditions determine the columns $W_\beta$ with $\beta >
2\Nin$ up to a unitary transformation on these columns alone, and
likewise the remaining columns are determined up to a unitary matrix
which combines them.

\subsection{Density matrix and entropy}

We can write $\calh = \tilde\calh \otimes \calg$ where $\tilde
\calh$\notationlabel{Hfree} is the Hilbert space of states of the
``free'' modes and $\calg$ is the Hilbert space of states of the
``ground state'' modes.  Let $|0\>_\gs\member\calg$ denote the ground
state of this system.  For any operator $A$ we can define an operator
$\tilde A$\notationlabel{tilde} that acts on $\tilde\calh$ via
$\<\tilde\alpha|\tilde A|\tilde\beta\> =
\<\tilde\alpha\otimes 0_\gs| A | \tilde\beta\otimes 0_\gs\>$ for all
$\tilde\alpha,\tilde\beta\member\tilde\calh$.

Let $\rho$ describe a vacuum-bounded state and let $\rho_\gs =
\Tr_\free\rho$.  Then $\Tr\rho_\gs\bdag_\beta b_\beta = 0$.  This
defines the vacuum state, so $\rho_\gs = |0\>_\gs\<0|_\gs$ and thus
 \be
\rho = \rhofree \otimes\left(|0\>_\gs\<0|_\gs\right)\,.
\ee
If we write $\rhofree$ in diagonal form,
\be
\rhofree = \sum_\alpha P_\alpha |\tilde\alpha\>\<\tilde\alpha|\,,
\ee
then
\be
\rho = \sum_\alpha P_\alpha |\tilde\alpha\otimes 0_\gs\>
\<\tilde\alpha \otimes 0_\gs|\,.
\ee
The entropy is
\be
S = -\Tr\rho\ln\rho = - \sum_\alpha s(P_\alpha) = - \Tr\rhofree\ln\rhofree\,.
\ee
Since any vacuum-bounded $\rho$ has this form, a variation of $\rho$
that preserves the constraints must also have this form,
\be
\ddelta\rho = \drhofree \otimes \left(|0\>_\gs\<0|_\gs\right)\,.
\ee
If $S = -\Tr(\rho + t\ddelta\rho) \ln(\rho+t\ddelta\rho)
= -\Tr(\rhofree + t\drhofree) \ln(\rhofree+t\drhofree)$, then
\be
{dS\over dt} = - \Tr\drhofree\ln\rhofree\,.
\ee
If $\rho$ maximizes $S$ subject to the constraints, then we must have
$\Tr\drhofree\ln\rhofree$ for any variation $\drhofree$ that
preserves the constraints, i.e.\ for which $\Tr\drhofree = 0$ and
\be
\Tr \ddelta\rho H = \Tr \ddelta\rho x_i x_j = \Tr  \ddelta\rho P_i P_j = 0\,.
\ee
Now for any $A$, $\Tr\ddelta\rho A = \Tr (\drhofree\otimes(|0_\gs\>\<0_\gs|))A$.
When we take the trace we only need to sum over states of the form
$|\tilde\alpha\otimes 0_\gs\>$.  Thus
\be[opistilde]
\<A\> = \Tr\ddelta\rho A = \sum_{\tilde\alpha\tilde\beta}
\<\tilde\alpha|\drhofree|\tilde\beta\>
\<\tilde\beta\otimes 0_\gs|A|\tilde\alpha\otimes 0_\gs\>
= \Tr\drhofree\tilde A = \<\tilde A\>\,.
\notationlabel{tildeH}
\ee

Thus we are looking for $\rhofree$ that maximizes
$S=-\Tr\rhofree\ln\rhofree$ subject to the constraints
$\Tr\drhofree\tilde H = \Tr \drhofree \widetilde{x_i x_j} = \Tr
\drhofree\widetilde{P_i P_j} = 0$, where $i$ and $j$ range over the
outside oscillators.  There is no longer a problem of zero eigenvalues
of $\tilde\rho$.

From \eqref{rhoform} we expect $\rhofree$ to have the form
\be[ourrhoform]
\rhofree \propto e^{-\beta\left(\tilde H + f_{ij}\widetilde{x_i x_j}
+g_{ij}\widetilde{P_i P_j}\right)}\,.
\ee
We can write this in a more familiar way as
\be
\rho \propto e^{-\beta H'}\,,
\ee
where \usenotation{H'} is a fictitious Hamiltonian for these oscillators,
\be[Hprimeform]
H' = \tilde H + f_{ij}\widetilde{x_i x_j} +g_{ij}\widetilde{P_i P_j}\,.
\notationlabel{fg}
\ee

\subsection{New coordinates}
\label{sec:startreduction}

We would now like to introduce new coordinates $\bw$ as follows: The
first $\Nin$ $\bw$ coordinates will be the inside oscillator
coordinates, $\bw_\in = \bx_\in$.  The last $N_\gs$ coordinates are
the ground state normal modes, $\bw_\gs = \by_\gs$.  The remaining
$\Nin$ coordinates can be any coordinates that are independent of
those specified so far; we will make a particular choice later.

To do this, we proceed as follows:  From \eqsref{Yform} and
(\ref{eqn:Yinvform}) we have

\be
Y Y^{-1} = 
\splitmatrix(Y_\infree Y^{-1}_\freein, Y_\infree Y^{-1}_\freeout,
 Y_\outfree Y^{-1}_\freein,
 Y_\outfree Y^{-1}_\freeout+ Y_\outgs Y^{-1}_\gsout)
= \sizedsplitmatrix(I, 0, 0, I)(\Nin, \Nout)(\Nin, \Nout)
\ee
and
\be
Y^{-1} Y =
\splitmatrix(Y^{-1}_\freein Y_\infree+Y^{-1}_\freeout Y_\outfree,
	Y^{-1}_\freeout Y_\outgs,Y^{-1}_\gsout Y_\outfree,
	Y^{-1}_\gsout Y_\outgs)
= \sizedsplitmatrix(I, 0, 0, I)(\Nfree, \Ngs)(\Nfree, \Ngs)\,.
\ee
In particular, $Y^{-1}_\gsout Y_\outgs = I$.  We would like to extend
$Y^{-1}_\gsout$ and $Y_\outgs$ into square matrices $R$ and $R^{-1}$ with
\be
\notationlabel{R}
R = \sizedjoinmatrix(D, Y_\outgs)(\Nin, \Ngs)(\Nout)
\ee
and
\be
R^{-1} = \sizedstackmatrix(D', Y^{-1}_\gsout)(\Nout)(\Nin,\Ngs)
\ee
This means that we must find $D$ and $D'$\notationlabel{DDprime} such that
\blea[dconstraints]
Y^{-1}_\gsout D & = & 0\\
D' Y_\outgs & = & 0\\
D' D & = & I\,.
\elea
There are many possible choices of $D$ and $D'$ that satisfy
\eqsref{dconstraints}.  Here we proceed as follows:  Let
\blea
Z & = & \stackmatrix(Z_\in, Z_\out)\\
Z^{-1} & = & \joinmatrix(Z^{-1}_\in, Z^{-1}_\out)
\elea
so that $Z_\in Z^{-1}_\in = I$, $Z_\out Z^{-1}_\out = I$ and
$Z^{-1}_\in Z_\in + Z^{-1}_\out Z_\out = I$.
Now let
\blea[ddbardef]
\bar D & = & \half Z_\out Z_\in^T = \<0|\xop|0\>_\outin\\
\bar D' & = & \half Z_\in^\invt Z^{-1}_\out = \<0|\pop|0\>_\inout\,,
\notationlabel{DDbar}
\elea
let $A$ and $B$
be $\Nin \times \Nin$ matrices with 
\be[abinv]
A B = (\bar D' \bar D)^{-1}
\ee
and let
\blea[ddprime]
D & = & \bar D A\\
D' & = & B \bar D'\,.
\elea
We also divide
\be
W = \joinmatrix(W_\free, W_\gs)\,.
\ee
Since $Y=ZW$ and $Y^{-1} = W^T Z^{-1}$, we have
\blea[zw-zero]
Z_\in W_\gs &= & 0\\
W_\gs^T Z^{-1}_\in & = & 0\,.
\elea
Thus
\be
D'Y_\outgs \propto B Z_\in^\invt Z^{-1}_\out Z_\out W_\gs
= B Z_\in^\invt W_\gs - B Z_\in^\invt Z^{-1}_\in Z_\in W_\gs
= 0
\ee
by \eqsref{zw-zero} and their transposes.  Similarly
\be
Y^{-1}_\gsout D \propto W_\gs^T Z^{-1}_\out Z_\out Z_\in^T A
= W_\gs^T Z_\in^T A - W_\gs^T Z^{-1}_\in Z_\in Z_\in^T A = 0\,.
\ee
From \eqref{abinv} we find $D'D = I$.  Thus the matrices $D$ and $D'$
satisfy \eqsref{dconstraints}.  We still have the freedom of
choosing the matrix $A$ arbitrarily.

Now let
\be
Q = \sizedsplitmatrix(I,0,0,R)(\Nin,\Nout)(\Nin,\Nout)
\notationlabel{Q}
\ee
and define $\bw$ by $\bx = Q\bw$ so $\Px = Q^\invt \Pw$.  Then
$\bw_\in = \bx_\in$ and $\bw_\gs = \by_\gs$ as desired.

\label{sec:xfromw}

\subsection{Reduced operators}

We would like to recast our problem in terms of $\bw_\free$, the the
first $\Nfree$ $\bw$ coordinates.  First we look at the operators $x_\mu
x_\nu = \xopsub_{\mu\nu}$ and $P_\mu P_\nu = \popsub_{\mu\nu}$.  If we write
\blea
Q &=& \joinmatrix(Q_\free, Q_\gs)\\
Q^{-1} &=&  \stackmatrix(Q^{-1}_\free, Q^{-1}_\gs)\,,
\elea
we have $\bx = Q \bw
= Q_\free \bw_\free + Q_\gs \bw_\gs$ and
$\bP = Q^\invt \Pw = Q^\invt_\free (\Pw)_\free + Q^{-1}_\gs
(\Pw)_\gs$, so
\blea
\xop &=& Q_\free \wopsub_\freefree Q_\free^T + Q_\gs \wopsub_\gsgs
Q_\gs^T\\
\pop &=& Q^\invt_\free \pwopsub_\freefree Q^{-1}_\free
+ Q^\invt_\gs  \pwopsub_\gsgs Q^{-1}_\gs\,.
\elea
Now we form the reduced operators $\widetilde{x_\mu x_\nu}$ and
$\widetilde{P_\mu P_\nu}$.  Since $\bw_\gs = \by_\gs$ and the $\by_\gs$
modes are in the ground state by definition, $\widetilde\wopsub_\gsgs =
\widetilde\pwopsub_\gsgs = (1/2) I$, and so
\blea[expandoptilde]
\tilde\xop &=& Q_\free \wopsub_\freefree Q_\free^T + \half Q_\gs Q_\gs^T\\
\tilde\pop &=& Q^\invt_\free \pwopsub_\freefree Q^{-1}_\free
+ \half Q^\invt_\gs Q^{-1}_\gs\,.
\elea
In each case the second term is just a constant.
Now
\be
H = \half\left(P_\mu P_\mu + K_{\mu\nu} x_\mu x_\nu\right)
= \half\Tr\left(\pop + K\xop\right)
\ee
where the trace is over the oscillator indices. Thus
\bea[tildeHdef]
\notationlabel{tildeH}
\tilde H & = & \half\Tr\left(Q^\invt_\free \pwopsub_\freefree Q^{-1}_\free
+ K Q_\free \wopsub_\freefree Q_\free^T\right)+\text{const}\nonumber\\
& = & \half\Tr\left(\tilde T \pwopsub_\freefree + \tilde K
\wopsub_\freefree \right) + \text{const}
\eea
where
\blea
\notationlabel{tildeK}\notationlabel{tildeT}
\tilde T &=& Q^{-1}_\free Q^\invt_\free = 
\sizedsplitmatrix(I, 0, 0, D' D'^T)(\Nin,\Nin)(\Nin,\Nin)\\
\tilde K &=& Q_\free^T K Q_\free =
   \sizedsplitmatrix(K_\inin, K_\inout D, D^T K_\outin, D^T K_\outout D)(\Nin,\Nin)(\Nin,\Nin)
\label{eqn:ktildedef}\,.
\elea

The constant term in \eqref{tildeHdef} is
\be
\half\Tr\left(K Q_\gs Q_\gs^T+Q^\invt_\gs Q^{-1}_\gs\right)
= \half\Tr\left(Y_\outgs^T K_\outout Y_\outgs + Y^\invt_\gsout
Y^{-1}_\gsout\right)\,.
\ee
It depends on the ground state modes only and is just part of
the zero-point energy.  It will be the same in the vacuum and in a
vacuum-bounded state.  Thus if instead of \eqref{tildeHdef} we use
\be[tildeHdef-noconst]
\tilde H = \half\Tr\left(\tilde T \pwopsub_\freefree + \tilde K
\wopsub_\freefree \right)
\ee
we are just shifting $\tilde H$ by a constant term and thus changing
the zero-point energy.

Now we would like to make this reduced system look as much as possible
like the system we started with.  Let $B_1$ be some matrix such that
$B_1^T B_1 = (\bar D' \bar D'^T)^{-1}$, let $B_2$ be a unitary matrix
to be determined, and let $B = B_2 B_1$.  Then $D' D'^T = I$, so
$\tilde T = I$.  This gives $A = A_1 A_2$ where $A_1 = (\bar D' \bar
D)^{-1} B_1^{-1}$ and $A_2 = B_2^T$.  Let $D_1 = \bar D A_1$ and let
\be
\tilde K_1= \splitmatrix(K_\inin, K_\inout D_1, D_1^T K_\outin, D_1^T K_\outout D_1)\,.
\ee

Now
\be
K_\inout = \lowerleftnonzero{-g}\,,
\ee
so $K_\inout D_1$ is nonzero only in the last row.  Thus the last
$\Nin+1$ rows and columns of $\tilde K_1$ look like
\be
\sizedsplitmatrixul(2g, ?, ?, ?)(1,\Nin)(1,\Nin)\,.
\ee
A matrix of size $(\Ninn)\times(\Ninn)$ can be put in tridiagonal form by
a unitary transform of the form
\be
\sizedsplitmatrixul(1, 0, 0, U)(1,\Nin)(1,\Nin)
\ee
which can be constructed using the Householder process.  (See, e.g.,
\cite{numrecip:book} section 11.2.)
We will use this to choose $A_1 = U$, so that $\tilde K$ will be
tridiagonal.  For each off-diagonal element in the resulting
tridiagonal matrix there is a choice of sign, and we will choose them
all to be negative.  Thus in the $\bw$ coordinates, each oscillator
has unit mass and is coupled only to its neighbors.  This
completely specifies the matrices $A$ and $B$ and thus $D$ and $D'$.
Note that $D$ and $D'$ do not depend on the undetermined parts of $W$.

To make $H'$ in \eqref{Hprimeform} we can add to $\tilde H$ a
kinetic and potential term involving outside oscillators only.  The
potential term (disregarding a constant) is
\be
f_{ij} \widetilde{x_i x_j} = \Tr f \widetilde{\xopsub}_\outout = 
\Tr f Q_\outfree \wopsub_\freefree Q_\outfree^T\,.
\ee
Since
\be
Q_\outfree = \sizedjoinmatrix(0, D)(\Nin,\Nin)(\Nout)\,,
\ee
this term is equivalent
to adding an arbitrary term to just the lower right part of $\tilde
K$.  Similarly the kinetic term is 
\be
g_{ij} \widetilde{P_i P_j} = \Tr g \widetilde{\popsub}_\outout = 
\Tr g Q^\invt_\freeout \pwopsub_\freefree Q^{-1}_\freeout\,.
\ee
Since
\be
Q^{-1}_\freeout =  \sizedstackmatrix(0, D')(\Nout)(\Nin,\Nin)\,,
\ee
this term corresponds to adding an arbitrary term to just the
lower right part of $\tilde T$.

That is to say we can write
\be[hprimeform]
H' = \half\Tr\left(T'\pwopsub_\freefree + K'\wopsub_\freefree\right)
\ee
with
\blea[tkform]
\notationlabel{T'}\notationlabel{K'}
T' & = & {\arraycolsep 5pt \splitmatrix(I, 0, 0, ?)}\\
K' & = & \splitmatrix(K_\inin, K_\inout D, D^T K_\outin, ?)\,.
\elea

\subsection{Reduced constraints}

Now we rewrite our constraints, \eqsref{rhoconstraints}, in terms of the
$\bw$ coordinates.  We will keep only the parts of the constraint
equations that are not automatically satisfied because of the
ground-state modes.  For the expectation value constraints, 
from \eqsref{opistilde} and (\ref{eqn:expandoptilde}) we have
\blea
\<\xop\>_\outout & = & 
\widetilde{\xopexsub}_\outout = Q_\outfree\<\wop\>_\freefree Q_\outfree^T
+ \half Q_\outgs Q_\outgs^T\nonumber\\
&=& D \<\wop\>_\midmid D^T + \text{const}\\
\<\pop\>_\outout & = & \widetilde{\popexsub}_\outout
= {Q^{-1}_\freeout}^T \<\pwop\>_\freefree Q^{-1}_\freeout
+ \half Q^\invt_\gsout Q^{-1}_\gsout\nonumber\\
& = & D'^T \<\pwop\>_\midmid D' + \text{const}
\elea
where $\bw_\mid$ means the outside elements of $\bw_\free$, i.e.\
$w_{\Nin+1}\ldots w_{2\Nin}$.

These expectation value matrices must be the same in the vacuum as in
the vacuum-bounded state.  We can accomplish this by requiring that
$\<\pwop\>_\midmid$ and $\<\wop\>_\midmid$ are the same as in the
vacuum.  Thus we have reduced the problem to one that has only
$\Nin(\Nin+1)$ expectation value constraints, regardless of the value
of $\Nout$.

For the energy constraint, \eqref{Hrhoconstraint}, we are concerned
only with the renormalized energy $\Tr\rho H - \<0|H|0\>$.  Thus the
constant term in \eqref{tildeHdef} does not contribute, and we can
use $\tilde H$ from \eqref{tildeHdef-noconst}.  Once again there is no
dependence on $\Nout$.

\subsection{Derivation based on inside functions}

We would now like to redo the proceeding calculation in a way which
does not depend on the number of outside oscillators.  Then we can
remove the box from our system by taking $L \rightarrow\infty$ and $N
\rightarrow \infty$.  It appears that we have used the matrices $D$
and $D'$ which have an index that runs from $1$ to $\Nout$.  However,
we have used them only in particular combinations.  The quantities
which we need in our calculation are
\begin{enumerate}\parskip 0pt
\item $K_\inout \bar D$
\item $\bar D'\bar D$
\item $\bar D'\bar D'^T$
\item $\bar D^T K_\outout \bar D$.
\end{enumerate}
Each of these quantities is an $\Nin\times\Nin$ matrix, so it is
reasonable to imagine that they do not depend on $\Nout$ in the
$N\rightarrow\infty$ limit.

We proceed as follows:  From \eqsref{ddbardef} we have
\blea
\bar D &=& \xopexsub_\outin\\
\bar D' &=& \popexsub_\inout\,.
\elea
Any given element of $\xopex$ and $\popex$ has a smooth limit when
$\Nout$ is taken to infinity.  It is just a particular expectation
value of a half-line of coupled oscillators, which is a well-defined
problem.  We can express the above items in terms of such elements as
follows:
\begin{enumerate}
\item $K_\inout\bar D$ depends only on the first row of $\bar D$ which is
$\xopexsub_{\Ninn,\in}$ so it is already well-defined in the limit.

\item From \eqref{xoppinv} we have
\be[popxop]
\popex \xopex = \qtr I
\ee
so
$\bar D'\bar D = \popexsub_\inout \xopexsub_\outin = \qtr I -
\popexsub_\inin \xopexsub_\inin$, which does not depend on $\Nout$.

\item We expand $\popex \popex = \qtr Z^\invt Z^{-1} Z^\invt Z^{-1}$.  We
insert $I = Z \Omega_0 Z^T$ here to get
\be[kpop]
\popex \popex = \qtr Z^\invt \Omega_0 Z^{-1} = \qtr K\,.
\ee
Then $\bar D'\bar D'^T = \popexsub_\inout \popexsub_\outin
= \qtr K_\inin - \popexsub_\inin \popexsub_\inin$ which does not
depend on $\Nout$.

\item Using \eqsref{popxop} and (\ref{eqn:kpop}) we can write
\be
\xopex K = K \xopex = \popex \qquad\hbox{and }\qquad 
\xopex K \xopex = \qtr I
\ee
so
\bea
\qtr I &=& \xopexsub_\inout K_\outout \xopexsub_\outin
+ \xopexsub_\inin K_\inout \xopexsub_\outin\nonumber\\
&&+ \xopexsub_\inout K_\outin \xopexsub_\inin
+ \xopexsub_\inin K_\inin \xopexsub_\inin\nonumber\\
& = & \bar D^T K_\outout \bar D
+ \xopexsub_\inin (\popexsub_\inin - K_\inin \xopexsub_\inin)\nonumber\\
&&+ (\popexsub_\inin - \xopexsub_\inin K_\inin) \xopexsub_\inin
+ \xopexsub_\inin K_\inin \xopexsub_\inin
\eea
and so
\be
\bar D^T K_\outout \bar D = \qtr I
- \xopexsub_\inin \popexsub_\inin
- \popexsub_\inin \xopexsub_\inin
+ \xopexsub_\inin K_\inin \xopexsub_\inin
\ee
which does not depend on $\Nout$.

\end{enumerate}

Thus we can now take $N\rightarrow\infty$ with $\Nin$ fixed and have a
well-defined problem in terms of $\tilde K$ with a finite number of
free parameters.

\label{sec:endfirstreduction}

\subsection{Calculation of the reduced vacuum}

We are trying to compute $\tilde K$ in \eqref{ktildedef}
in the $\Nout\rightarrow\infty$ limit.  We will keep $\Nin$ and
the oscillator spacing $L_1 \equiv L/(N+1)$ fixed.
With the normalization in \eqref{Znorm} the vacuum normal mode matrix is
given by
\be
Z_{\mu\nu} = \sqrt{2\over N+1}{\sin k_\nu \mu\over\sqrt{\omega_\nu}}
\ee
with
\be
k_\nu = {\pi \nu\over N+1}
\ee
and
\be
\omega_\nu = {2 (N+1)\over L} \sin {k_\nu\over 2}
= {2 \over L_1} \sin {k_\nu\over 2}\,.
\ee
Thus
\bea
\xopexsub_{\mu\nu} & = & \half\left(Z Z^T\right)_{\mu\nu}
= {1\over N+1}\sum_{\alpha=1}^N {\sin k_\alpha \mu \sin k_\alpha \nu \over \omega_\alpha}\nonumber\\
& = & {L_1\over 2(N+1)} \sum_{\alpha=1}^N {\sin k_\alpha \mu \sin k_\alpha \nu\over \sin
(k_\alpha/ 2)}\,.
\eea
Now we use
\be[cosdif]
\cos(\theta-\phi) - \cos(\theta+\phi) = 2 \sin\theta \sin \phi
\ee
to write
\be[xopsum]
\xopexsub_{\mu\nu} = {L_1\over 4(N+1)}\sum_{\alpha=1}^N
   {\cos k_\alpha (\mu-\nu) - \cos k_\alpha(\mu+\nu) \over \sin(k_\alpha/2)}\,.
\ee
Using \eqref{cosdif} again, for any number $a$ we can write
\be
{\cos k_\alpha (a-1) - \cos k_\alpha a\over \sin(k_\alpha/2)} = 2 \sin k_\alpha (a-1/2)
\ee
and thus
\be[cossum]
{\cos k_\alpha (\mu-\nu) - \cos k_\alpha(\mu+\nu) \over \sin(k_\alpha/2)}
= 2 \sum_{a=\mu-\nu+1}^{\mu+\nu} \sin k_\alpha (a-1/2)\,.
\ee
If we put \eqref{cossum} into \eqref{xopsum} and bring the sum over $\alpha$
inside the sum over $a$ we get a sum that we can do,
\be
2 \sum_{\alpha=1}^N \sin {\pi \alpha (a-1/2) \over N+1}
= \cot {\pi(2a-1)\over 4(N+1)} + (-1)^a\,.
\ee
Now we sum this over $a$.  Since $\mu-\nu$ and $\mu+\nu$ have the same parity,
the $(-1)^a$ term does not contribute and we get
\be
\xopexsub_{\mu\nu} = {L_1\over 4(N+1)} \sum_{a=\mu-\nu+1}^{\mu+\nu} 
    \cot {\pi(2a-1)\over 4(N+1)}\,.
\ee

The sum over $N$ has been eliminated.  In the $N\rightarrow\infty$
limit, the argument of $\cot$ goes to zero and so we can
use $\cot x = 1/x + O(x^{-3})$ to get
\be
\xopexsub_{\mu\nu} = {L_1 \over \pi} \sum_{a=\mu-\nu+1}^{\mu+\nu} {1\over 2a-1}\,.
\ee
The sum can be done using special functions:
\be[finalxop]
\xopexsub_{\mu\nu} = {L_1 \over 2\pi}
\left[\psi\left(\mu+\nu+\half\right) - \psi\left(\mu-\nu+\half\right)\right]
\notationlabel{digamma}
\ee
where $\psi$ is the digamma function $\psi(x) = \Gamma'(x)/\Gamma(x)$.

\label{sec:xopexform}

To compute $\popex$ we write the inverse of the normal mode matrix,
\be
Z^{-1}_{\mu\nu} = \sqrt{2\omega_\mu\over N+1}\sin k_\mu \nu
\ee
and
\bea
\popexsub_{\mu\nu} & = & \half\left(Z^\invt Z^{-1}\right)_{\mu\nu}
 = {1\over N+1}\sum_{\alpha=1}^N\omega_\alpha\sin k_\alpha \mu \sin k_\alpha \nu\nonumber\\
& = & {2\over L_1(N+1)} \sum_{\alpha=1}^N\sin {k_\alpha\over 2}\sin k_\alpha \mu \sin k_\alpha \nu\,.
\eea
In this case the sum can be done directly.  Using \eqref{cosdif} we can write
\be
\popexsub_{\mu\nu} = \popone_{\mu-\nu} - \popone_{\mu+\nu}
\ee
where
\bea
\popone_\lambda & \equiv & {1\over L_1(N+1)} \sum_{\alpha=1}^N \sin {k_\alpha\over 2} \cos k_\alpha \lambda\nonumber\\
& = & {1\over 4 L_1 (N+1)}
\left[\cot{\pi(2\lambda+1)\over 4(N+1)}
 - \cot{\pi(2\lambda-1)\over 4(N+1)} - 2 (-1)^\lambda\right]\,.
\eea
Again we take $N \gg \lambda$ to get
\be
\popone_\lambda = {1\over \pi L_1}\left({1\over 2\lambda+1}
 - {1\over 2\lambda-1}\right)\\
= - {2\over \pi L_1(4\lambda^2-1)}
\ee
and
\be[finalpop]
\popexsub_{\mu\nu} = {2\over \pi L_1}
\left({1\over 4(\mu+\nu)^2-1} - {1\over 4(\mu-\nu)^2-1}\right)\,.
\ee
\Eqsref{finalxop} and (\ref{eqn:finalpop}) give $\xopex$ and $\popex$
in the $\Nout\rightarrow\infty$ limit.  Using these values in the
procedure of sections \ref{sec:startreduction} through
\ref{sec:endfirstreduction} we can compute the matrix $\tilde K$
numerically for a system with inside length $\Lin$ but no outside box.

\label{sec:endreduction} 

\section{Calculation of the vacuum-bounded state}

Once we have computed $\tilde K$ we can go on to look for a
vacuum-bounded state.  We are working entirely with the reduced
coordinates.  With
\be[solvingrhoform]
\rho\propto e^{-\beta H'}
\ee
and $H'$ as in Eqs. (\ref{eqn:hprimeform}--\ref{eqn:tkform}) we have
one number, $\beta$, and two symmetric $\Nin\times\Nin$ matrices,
$K'_\midmid$ and $T'_\midmid$, that we can adjust.  The constraints
(\ref{eqn:rhoconstraints}) involve one scalar constraint, for total
energy, and two $\Nin\times\Nin$ symmetric matrices of constraints,
for $w_m w_n$ and $P_{w_m} P_{w_n}$.  There are equal numbers of
equations to satisfy and free parameters to adjust, and so, if we are
lucky, we will be able to find a solution.  If we do find a solution,
we know it is unique from the arguments of \secref{unique}.

\subsection{Expectation values}
\label{sec:exptvals}

To actually solve these equations we will need to compute the
expectation values of $w_m w_n$ and $P_{w_m} P_{w_n}$ given the
density matrix (\ref{eqn:solvingrhoform}).  To do this we find the
normal modes of $H'$.  The Hamiltonian $H'$ gives rise to classical
equations of motion
\be
{d^2 \bw_\free\over dt^2} = - T' K' \bw_\free\,,
\ee
so we look for eigenvectors $\walpha_\free$ that satisfy
\be
T' K' \walpha_\free = \omega_\alpha^2 \walpha_\free\,.
\ee

The eigenvectors will be complete, so that we can define new
coordinates $u_\alpha$ via $\bw_\free = \sum u_\alpha \walpha_\free$, which
will then obey the equations of motion
\be
{d^2u_\alpha\over dt^2} = - \omega_\alpha^2 u_\alpha\,.
\ee
We can choose the norms of the eigenvectors so that\footnote{This is a
different normalization than we used for $Z$ in
\secref{oldgroundstate}.  In the present normalization the
original vacuum modes would all have the same amplitude.  In
\secref{oldgroundstate} the modes were multiplied by a factor
of $\omega_\alpha^{-1/2}$ relative to the convention here.}
$\walpha_\free\cdot K'\bw^\beta_\free = \omega_\alpha^2 \delta_{\alpha\beta}$ 
and group the eigenvectors into a matrix \usenotation{U} via
$U_{\mu\alpha} = w^\alpha_\mu$.
Then we will find that
$T'_{\mu\nu} = w^\alpha_\mu w^\alpha_\nu$, or
\be
K' = U^\invt \Omega^2 U^{-1} \qquad\text{and}\qquad T' = U U^T 
\ee
where $\Omega_{\alpha\beta} = \omega_\alpha \delta_{\alpha\beta}$.
We can then substitute
\be
\bw_\free = U \bu\qquad (\Pw)_\free = U^\invt \Pu
\ee
into $H'$ to get
\bea
H' &=& \half\left(\Pu\cdot\Pu+\bu\cdot\Omega^2\bu\right)\nonumber\\
&=& \half\sum_\alpha({P_{u_\alpha}}^2+\omega_\alpha^2 u_\alpha^2) 
\equiv \sum_\alpha H'_\alpha\,.
\eea
This is the Hamiltonian for a set of (fictitious) uncoupled
oscillators with frequencies $\omega_\alpha$.  The expectation values of
the $u$ and $P_u$ are easily computed,
\blea
\<u_\alpha u_\beta\> & = &
{1\over 2\omega_\alpha} \delta_{\alpha\beta}\coth{\beta\over 2 \omega_\alpha}\\
\<P_{u_\alpha} P_{u_\beta}\> & = &
{\omega_\alpha\over 2} \delta_{\alpha\beta}\coth{\beta\over 2 \omega_\alpha}
\elea
so that in terms of the $\bw$ coordinates we have
\blea[wexptvals]
\<w_m w_n\> & = & \sum_\alpha{1\over 2\omega_\alpha}U_{m\alpha} U_{n\alpha}\coth{\beta\over2}\omega_\alpha\label{eqn:wwexptval}\\
\<P_{w_m} P_{w_n}\> & = & \sum_\alpha{\omega_\alpha\over 2}U^{-1}_{\alpha m} U^{-1}_{\alpha n}
\coth{\beta\over 2}\omega_\alpha\,.
\elea

In \secref{vb-computation} we will compute these expectation values
numerically in the vacuum state, given by $H' = \tilde H$ and $\beta =
\infty$, and require that they have the same values in the
vacuum-bounded state with finite $\beta$.

\section{The nature of the results}
\label{sec:understanding}

Before looking at the results of our computations, we would like to
learn as much as possible about the form that our answers must take.
We will take the frequencies $\omega_\alpha$ as given here and look at
the form of the normal modes $\walpha$.

The vectors $\walpha_\free$ satisfy the equation 
\be[eveqn]
T' K' \walpha_\free = \omega_\alpha^2 \walpha_\free
\ee
with
\be
T' = \splitmatrix(I, 0, 0, T')
\ee
and
\be
K' = \splitmatrix(K_\inin, K_\inout D, D^T K_\outin, K'_\midmid)\,.
\ee
However, we have tridiagonalized $\tilde K$, so only the lower left
element of $K_\inout D$ is nonzero.  In fact, in our numerical work we
will find that this element is always $-g$, just as it was in the
original $K_\inout$.  Thus we have 
{\arraycolsep 0pt
\be
K' = \left(\begin{array}{c|c}
K_\inin & \begin{array}{cc}
	 & \hspace{4pt} 0 \hspace{4pt} \\
	 \cline{1-1}
	 \mbox{\hspace{2pt} \small $-g$ \hspace{1pt}} & \vline\hfill\end{array} \\
\hline
\begin{array}{cc}
\hfill\vline & \mbox{\hspace{2pt} \small $-g$ \hspace{1pt}} \\
\cline{2-2}
\hspace{4pt} 0 \hspace{4pt} & \end{array} & \hspace{2pt}
K'_\midmid \end{array}\right)\,.
\ee
}

Writing out \eqref{eveqn} in components, we get
\be
T'_{\mu\nu} K'_{\nu\lambda} w^\alpha_\lambda = 
  \omega_\alpha^2 w^\alpha_\mu\,.
\ee
Taking only the inside components of the eigenvalue equation, we see
that
\be
K'_{a\mu} w^\alpha_\mu = \omega_\alpha^2 w^\alpha_a
\ee
for $a\le\Nin$.  That is to say,
\blea
2g w^\alpha_1 - g w^\alpha_2 & = & \omega_\alpha^2 w^\alpha_1 \\
-g w^\alpha_1 + 2g w^\alpha_2 - g w^\alpha_3 & = &
\omega_\alpha^2 w^\alpha_2 \\ 
& \cdots & \nonumber\\
-g w^\alpha_{\Nin-1} + 2 g w^\alpha_{\Nin} - g w^\alpha_\Ninn
& = & \omega_\alpha^2 w^\alpha_{\Nin}\,.
\elea
Taking $\omega_\alpha$ fixed, there are $\Nin$ equations involving $\Ninn$
unknown components of $\walpha$.  However, the equations are invariant under
a uniform rescaling of $\walpha$.  Thus these $\Nin$ equations fix
$w^\alpha_\mu$ for $\mu =1\ldots \Ninn$, except for
normalization.  The equations are readily solved, and the solution is
\be[definewin]
w^\alpha_\mu = N'_\alpha \sin \mu k'_\alpha\,,
\notationlabel{Nalpha}
\ee
where
\be
\cos k'_\alpha = 1 - {\omega_\alpha^2\over 2 g}\,,
\ee
and $N'_\alpha$ is an unknown normalization factor.
Here $k'_\alpha$ and $N'_\alpha$ can be complex, but
$\walpha$ must be real.  When $k'_\alpha$ is real we will write $k_\alpha =
k'_\alpha$ and $N_\alpha = N'_\alpha$ and call this a ``normal'' mode.
When $k'_\alpha$ is complex we can write
\blea
k'_\alpha & = & \pi + i k_\alpha\\
N'_\alpha & = & i N_\alpha\\
w^\alpha_\mu & = & (-)^{\mu-1} N_\alpha \sinh \mu k_\alpha
\elea
with $k_\alpha$ and $N_\alpha$ real.  We will refer to these as
``abnormal'' modes.

A similar calculation can be done for $U^{-1}$, the inverse of the
eigenvalue matrix $U$.  In this case we will find that
\be[defineuinv]
U^{-1}_{\alpha a} = N_\alpha'\sin a k'_\alpha \,.
\ee
\Eqref{defineuinv} has the same form as \eqref{definewin}, but applies only for
$a=1\ldots\Nin$.  That makes \eqref{defineuinv} less useful than
\eqref{definewin} for establishing a connection between the inside and
the outside region, and we will not use it further.

\addtocontents{toc}{\protect\pagebreak}
\chapter{Numerical Solution}

Two numerical calculations are necessary to solve the problem.  First we
must follow the procedure of sections
\ref{sec:startreduction}--\ref{sec:endreduction} to find the reduced
Hamiltonian $\tilde H$ and its vacuum state.  Then we must search for
parameter values $T'_\midmid$, $K'_\midmid$ and $\beta$ which produce a
state with a given energy $E_0$ but the same expectation values of the
outside oscillators.

In fact, we follow a slight variant of the above plan.  Instead of
fixing $E_0$, we hold $\beta$ fixed.  The resulting state has some
$E_0$ and solves the problem of maximizing the entropy for that $E_0$,
whatever it is.  By varying $\beta$ we can find states for various
energies.

In the end we do not use these results directly to calculate bounds on
the entropy.  Instead we derive a general principle from the numerical
calculations, and use this principle as an ansatz to derive a bound in
the next chapter.

\section{Numerical procedures}

Finding the reduced vacuum is a fairly straightforward problem in
numerical analysis.  The number of steps grows as $\Nin^3$.  However,
in order to produce accurate results for $\Nin \agt 6$ it is necessary to
use very high-precision floating-point numbers.  The necessary number
of bits of mantissa in the representation appears to be about 
$10\Nin$.

After finding the reduced vacuum we need to solve a set of
simultaneous nonlinear equations.  Such problems are in general quite
difficult, and require an iterative search for the correct parameter
values.  Here at least we know from \secref{unique} that there
cannot be more than one solution.  Although we have not been able to
prove that a solution always exists, in the numerical work we have
always been able to find one.

Once the search is sufficiently close to the correct answer it is
possible to use Newton's method, which converges quadratically, i.e.,
the number of correct digits doubles every step.  However, the basin
of attraction for Newton's method can be quite small and difficult to
find.  When Newton's method does not work, it is necessary to use some
other procedure to make progress toward the solution.  Here we used
the Powell hybrid method \cite{powell:method,powell:program}.  This
method moves in a direction which is a combination of the direction
Newton's method would suggest and the direction of steepest descent in
the mean square error of the function values.  Such a method has the
difficulty that it can get stuck at a local minimum of the mean square
error that is not a solution.  We have been lucky in that there do not
seem to be such local minima in this problem.\footnote{The current
formulation of the problem has $O(\Nin^2)$ parameters and equations.
It is possible to use the normal-mode frequencies and normalizations
as our parameters, which gives a problem with only $O(\Nin)$
parameters and equations.  However, this reduction loses the property
of having no local minima, and so in fact makes the problem harder to
solve.}

In the case that there are no local minima of the mean square error
in the function values, Powell's method is guaranteed to converge from
any starting point \cite{powell:method}.  However, it often converges
quite slowly for large systems, requiring many thousands of iterations
to make progress.  This has limited our numerical solutions to
problems with $\Nin \alt 15$.

The code was written in Lisp and executed on DEC$^{\text{TM}}$
Alpha$^{\text{TM}}$
workstations. All results presented here were computed using at least
38 decimal digits of precision.  The most precise calculation (the
reduced vacuum for $\Nin = 175$) used 636 digits.  The parameters
found in Powell's method reproduce the desired expectation values to
at least 17 significant digits.

We have made use of many routines from {\em Numerical Recipes}
\cite{numrecip:cdrom}.  However, the code that implements Powell's method was
written from scratch in Lisp following the outlines of
\cite{powell:method,powell:program} and has many extra features
including dynamic increase of working precision as the solution
converges.

\section{Reduction of the vacuum}

First we follow the procedures of sections
\ref{sec:startreduction}--\ref{sec:endreduction} to calculate the
matrix $\tilde K$ which gives the ground state of the $\bw_\free$
coordinates.  The result
is, of course, a tridiagonal matrix which gives a set of
self-couplings and nearest-neighbor couplings for the fictitious
oscillators $\bw_\free$.  We can express these couplings as multiples
of the couplings for a regular chain of $\Nfree$ oscillators with
spacing $L_1$.  Thus we write the self-coupling as
\be
\tilde K_{\mu\mu} = -g f_\mu\qquad \left(\notsum_\mu\right)
\ee
and the nearest-neighbor coupling as 
\be
\tilde K_{\mu,\mu+1} = 2g f_{\mu+1/2} \qquad \left(\notsum_\mu\right)\,.
\ee

These coupling coefficients converge rapidly to a universal form
\usenotation{f(x)} where $L_1 \mu\rightarrow x$ in the continuum limit.  Some
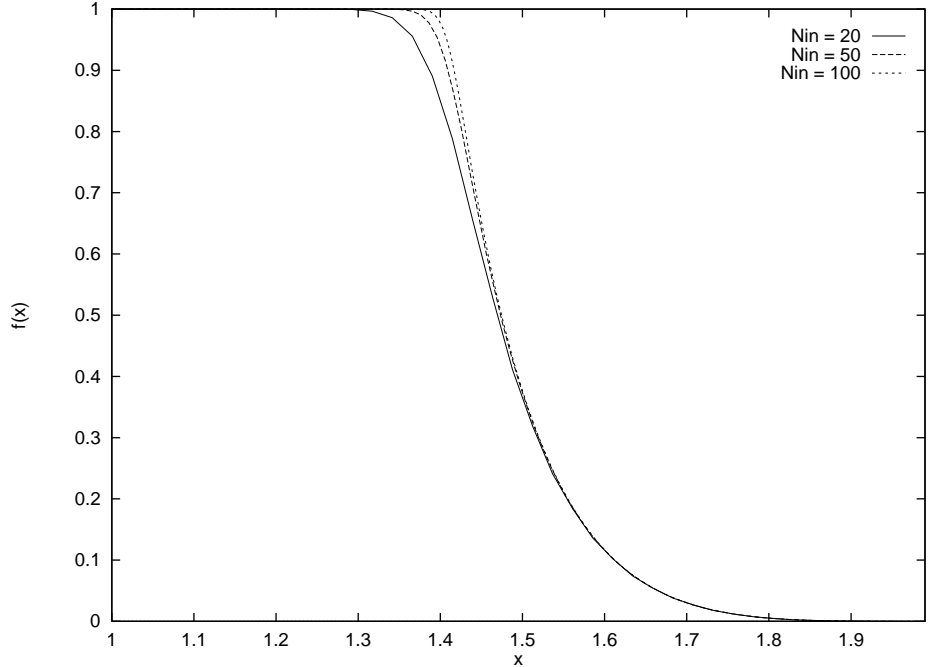
results are shown in \mypsfigr{fig4-1}{The ratio of the coupling
coefficients in the reduced problem to what they would be in a regular
problem of $\Nfree$ oscillators, plotted against $x = L_1\mu$, for
$\Lin = 1$}.  We can see that $f(x) \approx 1$ until $x\sim 1.4$ at
which point it begins to fall and asymptotically approaches $0$ as
$x\rightarrow 2$.  For values of $x$ near $2$, $f(x)$ is well fit by
\be[fxform]
f(x) = a (2-x)^4
\ee
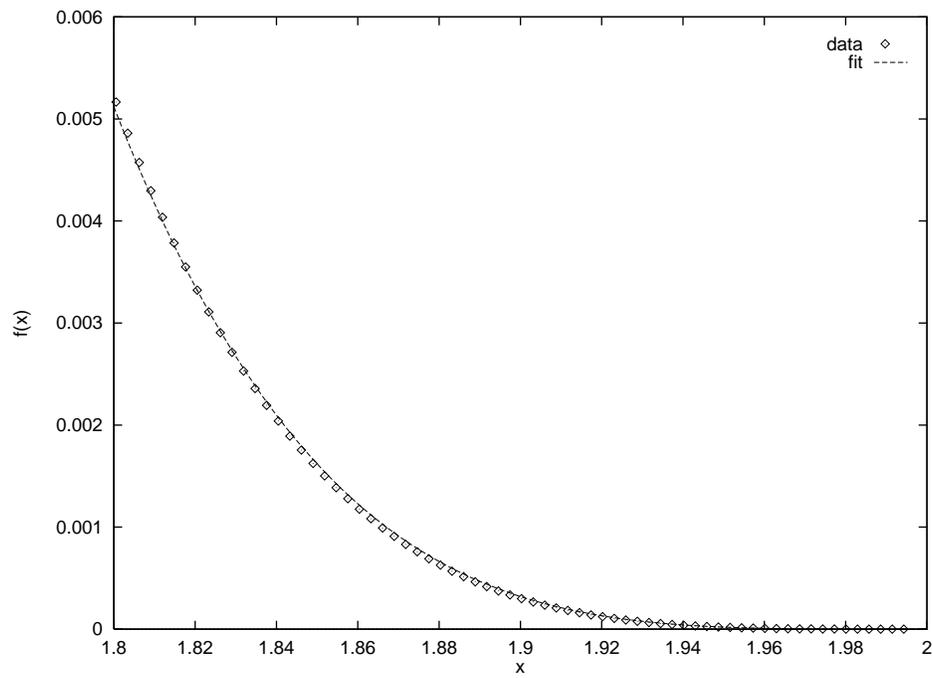
with $a\approx 3.2$, as shown in \mypsfigr{fig4-2}{The
coupling coefficient ratio for $\Nin=175$ in the region $x>1.8$ and
the fit $f(x) = 3.2 (2-x)^4$}.

\label{sec:smallfreqs}
Some typical normal modes of the reduced vacuum are shown in
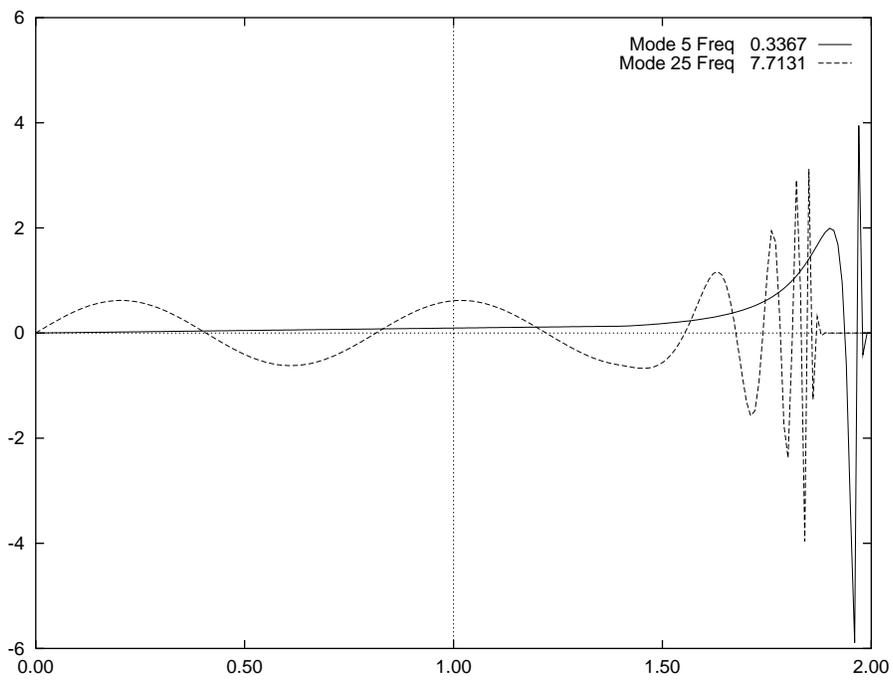
\mypsfigroff{fig4-3}{The $5^{\protect\text{th}}$ and
$25^{\protect\text{th}}$ normal modes in the reduced vacuum, computed
with $\Nin=175$}.  They
are sine waves while $f(x) \sim 1$ and then begin to oscillate faster
and faster as $f(x)$ shrinks.  At first the amplitude of the
oscillations grows but for larger $x$ it shrinks rapidly to zero.  The
wavenumbers in the inside region (and thus the frequencies) are
smaller than we would find for a rigid box because most of the
oscillations are in the part of the outside region where $f(x) \ll 1$.
In fact, as $\Nin\rightarrow\infty$ we would expect the low-lying
frequencies to go to zero, for the following reason.

We can find the frequencies by computing the normal modes of a Hamiltonian
\be
H = \int_0^L dx dy \left(T(x-y)\pi(x)\pi(y)+K(x-y)\phi(x)\phi(y)\right)\,,
\ee
which requires solving the eigenvector equation
\be[generaldiffeq]
\int_0^L dy dz T(x-y) K(y-z) g(z) = \lambda g(x)
\notationlabel{g(x)}
\ee
with the boundary conditions 
\blea[generalbcs]
g(0) &=& 0\\
g(L) &=& 0\label{eqn:rightbc}\,.
\elea
Since \eqref{generaldiffeq} is a second-order differential equation we
expect two degrees of freedom in the solution.  However, one degree of
freedom is manifestly the overall scale, which does not affect the
boundary conditions.  Since there are two boundary conditions but only
one free parameter, we can expect to find solutions only for
particular values of $\lambda$.  For example, for the usual scalar
field Hamiltonian the general solution to \eqref{generaldiffeq} would be
\be
g(x) = c \sin(\sqrt{\lambda} x + \delta)\,.
\ee
To satisfy \eqsref{generalbcs} we need to choose $\delta=0$ and
$\sqrt{\lambda} = n\pi/L$ for some integer $n$.

However, if we use
\be[redvacH]
H = \half \int_0^{2\Lin} f(x)\left[\pi(x)^2 
 + \left({d\phi\over dx}\right)^2\right] dx\,,
\ee
with $f(x)\rightarrow a(2-x)^4$, as suggested by
\figref{fig4-2}, we will get a continuum of frequencies.  The
problem is that since $f(x)\rightarrow 0$ as $x\rightarrow 2$ the
boundary condition there does not really constrain $g(x)$.  There can
be arbitrary changes in $g(x)$ near $x=2$ and so \eqref{rightbc}
can always be satisfied.  Since there is only one effective boundary
condition and one effective degree of freedom, we expect to be able
to find a solution for any $\lambda$.  Thus in the continuum limit
there are modes with arbitrarily low frequencies.  This is not an
unreasonable conclusion, since although the range of $x$ is finite, we
are using it to represent the infinite half-line.  In the infinite
vacuum there is no right-hand boundary condition, and
there are modes of every frequency.

This conclusion is confirmed by numerical results. In 
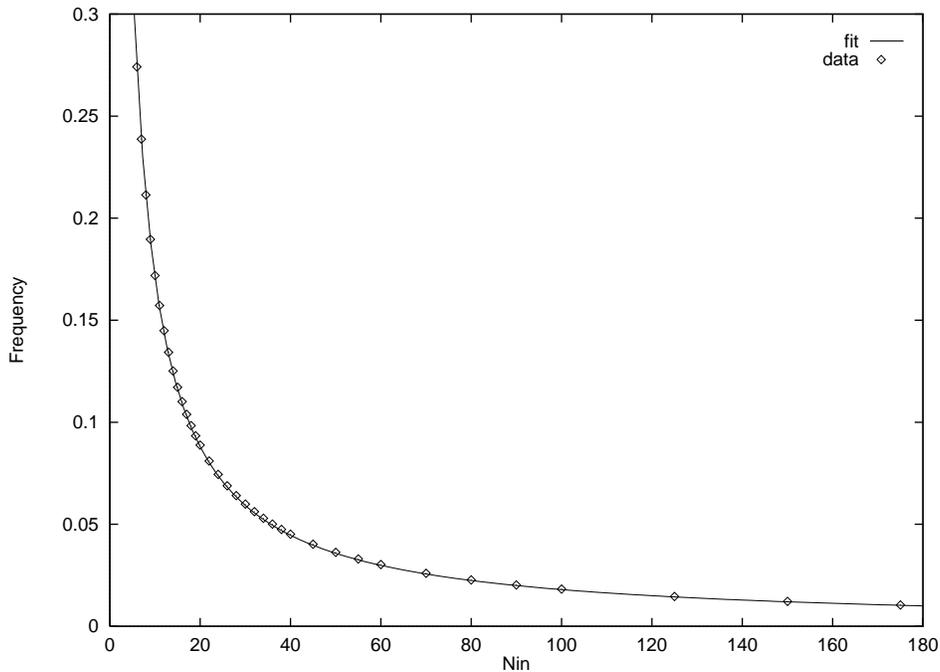
\mypsfigr{fig4-4}{The lowest frequency of the reduced
vacuum and the fit $1.81 x^{-1} - x^{-2}$} we plot
the lowest normal-mode frequency versus $\Nin$.  As shown in the
figure, the frequencies are well fit by a curve
\be
a x^{-1} - x^{-2}
\ee
with $a\approx 1.81$.  If this form is correct, in the $\Nin
\rightarrow\infty$ limit the lowest frequency goes to zero.

If we go to a vacuum-bounded state we will introduce some finite
temperature.  We then expect that the non-zero temperature will
increase the frequencies in such a way that there are only a finite
number of low-lying modes and thus a finite entropy.  However, in the
limit where $T\rightarrow 0$ we do expect an entropy-to-energy
relation equivalent to a system with infinitesimal frequencies.  This is
discussed in the appendix.

\section{Computation of the vacuum-bounded state}

\label{sec:vb-computation}
Once we have computed $\tilde K$ and the vacuum expectation values of
$w_m w_n$ and $P_{w_m}P_{w_n}$, we can proceed to look for the vacuum-bounded
state at a particular $\beta$.  To do this we vary the $\Nin(\Nin+1)$
independent components of $T'_\midmid$ and $K'_\midmid$ to find those which
reproduce the same expectation values of the $\Nin(\Nin+1)$ independent
operators $w_m w_n$ and $P_{w_m} P_{w_n}$, given in
\eqsref{wexptvals}.

To understand the numerical solution we look at the normal mode
frequencies and the forms of the normal modes.  When the mode is
``normal'' (i.e.\ real $k_\alpha'$ in \eqref{definewin}) there
is a sine wave in the inside region.  When the mode is
``abnormal'' there is essentially a growing exponential.  Typical modes
for a small number of oscillators are shown in
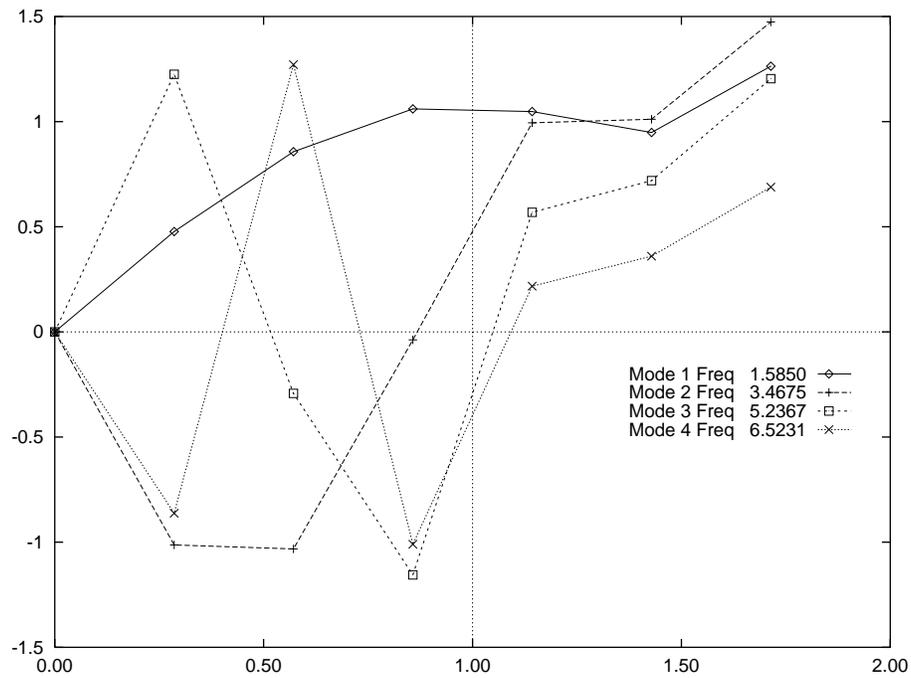
\mypsfigoff{fig4-5}{Modes and frequencies for the ``normal'' modes
of a system with $\Lin=1.0$, $\beta=2$, $\Nin=3$, $\Nfree=6$}
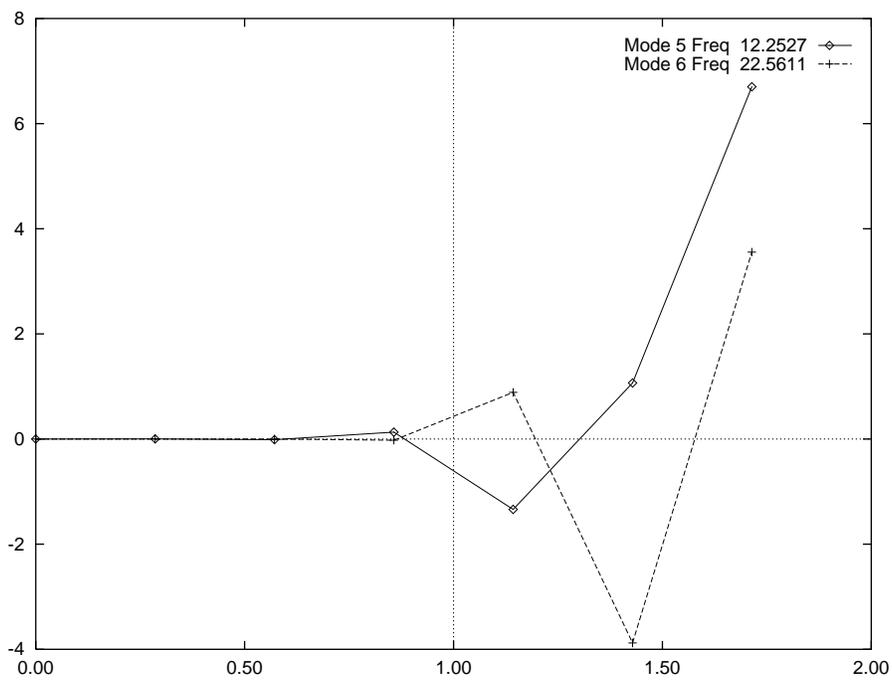
\mypsfigoff{fig4-6}{Modes and frequencies for the ``abnormal''
modes of a system with $\Lin=1.0$, $\beta=2$, $\Nin=3$,
$\Nfree=6$}\figsref{fig4-5} and \ref{fig:fig4-6}.

As $N$ becomes large, each ``normal'' mode and its frequency smoothly
approach a limit, providing that we use a normalization appropriate
for the continuum, which means that each mode must be rescaled by
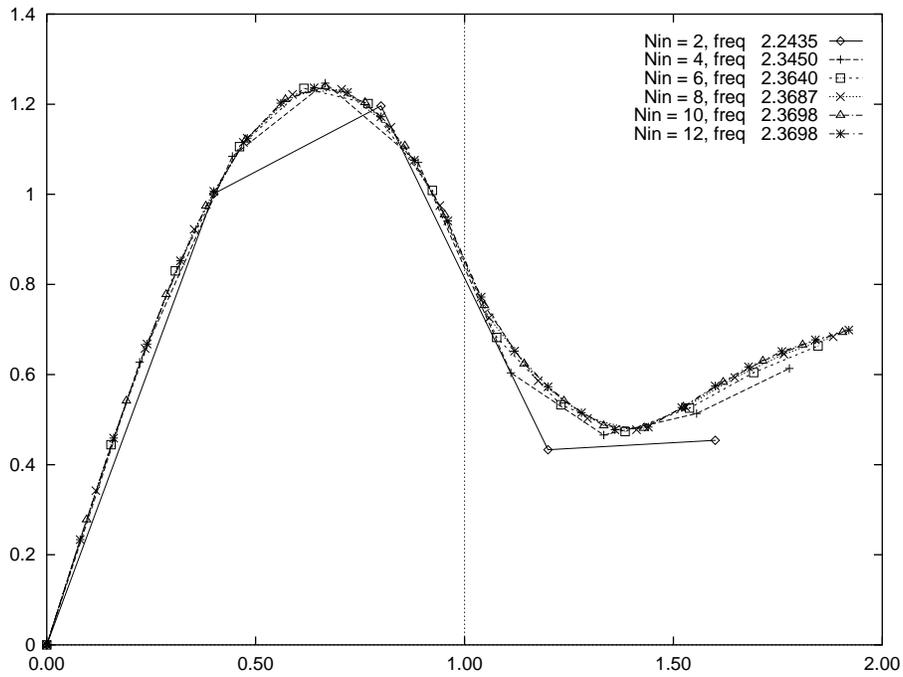
$L_1^{-1/2}$.  In \mypsfigroff{fig4-7}{The first normal mode,
rescaled to the continuum normalization, for various different numbers
of oscillators, with $\Lin=1.0$, $\beta=0.5$} we show the first
normal mode for various values of $\Nin$.  Note that this mode does
not appear to come down to zero at $x=2$.  As discussed in
\secref{smallfreqs}, this happens because $f(x)$ is going to $0$
at $x=2$ and so there is not really any coupling to the boundary at
that point.

As $N$ increases, each
abnormal mode and its frequency undergo a smooth evolution, until at
some point it disappears from the set of abnormal modes and is
replaced by a normal mode \label{sec:modeconvert} with very similar
form in the outside region, as shown in
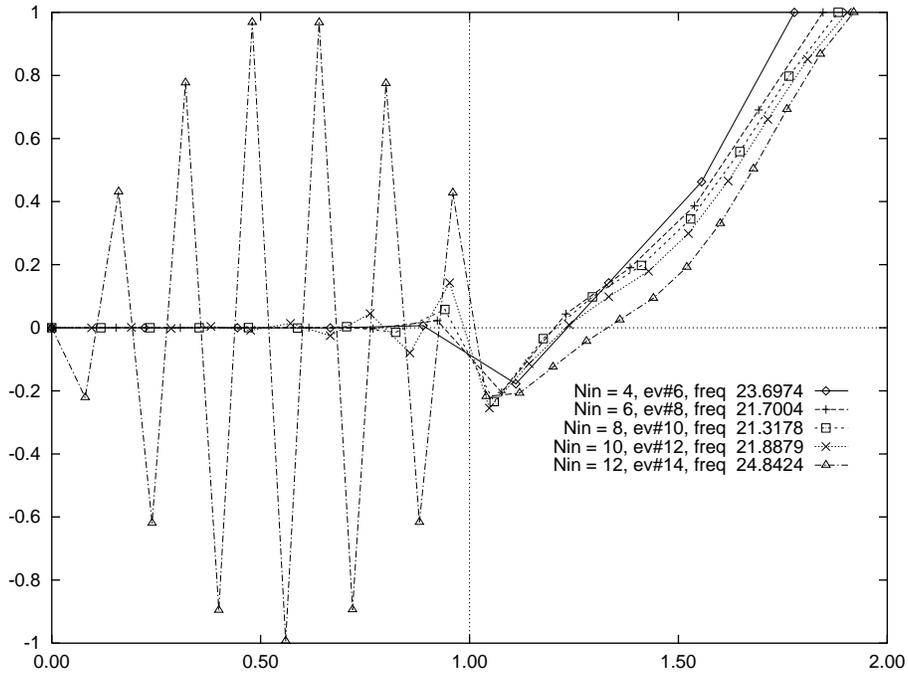
\mypsfigroff{fig4-8}{An abnormal mode for $\Lin=1.0$,
$\beta=0.5$ and various values of $\Nin$.  The normalizations are chosen
to be similar in the outside region. As $\Nin$ increases the mode
changes smoothly until at $\Nin=12$ there is no corresponding abnormal
mode, but instead one of the new normal modes has a very similar form
(but different normalization)}.  Because of this behavior, we believe
that if we could solve the continuum behavior directly we would find
just the ``normal'' modes.

\subsection{Evenly spaced wavenumbers}
\label{sec:wavenumber-spacing}

To address the problem directly for large energies would require
numerical solutions for large numbers of oscillators, which is
computationally intractable.  Instead we would like to extract from
the computations in the accessible regime a statement which will allow
us to extend our arguments to larger energies.  The most striking such
result is that the wavenumbers of the ``normal'' modes are nearly
evenly spaced.  The larger the energy, the more accurate is this
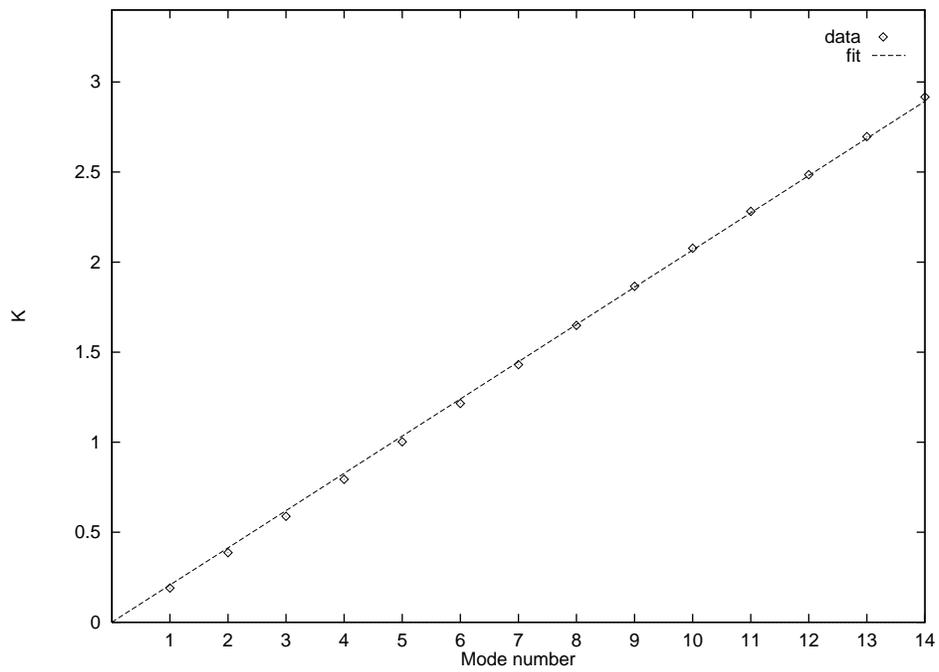
approximation.  In \mypsfigr{fig4-9}{The
numerically computed wavenumbers compared with the best-fit line
through the origin for $\Lin=1.0$, $\beta=0.5$, $\Nin=12$},
we show the wavenumbers for $\Lin=1.0$ and $\beta=0.5$, which
give an energy of about 0.094.  Even at this low energy the fit is
good to within a few percent of the typical wavenumber.  For larger
energies the points will lie correspondingly closer to the line.

Since we have set $N$ and $L$ to $\infty$, our problem has only
two dimensionful parameters, $\Lin$ and $E_0$.  Thus there is only one
dimensionless parameter, $\Lin E_0$, that characterizes the problem.
In the 3-dimensional black hole problem the equivalent parameter is $R
E_0$, which for the parameters in \eqref{goshwownumbers} is
about $10^{13}$. Thus for application to black holes we are interested
only in very large values of $\Lin E_0$, for which the linear
approximation for the ``normal'' modes will be very good.

Since the ``normal'' mode wavenumbers extend up to $k \sim \pi$, the
number of ``normal'' modes will be given an integer $\Nnorm \sim
\pi/k_1$ where $k_1$ is the interval between wavenumbers.  In addition
there will be $\Nfree - \Nnorm$ ``abnormal'' modes, with frequencies
$\omega_\alpha > 2/L_1$.  For $\Nin \gg \Lin/\beta$ these modes do not
contribute to the entropy, because they are exponentially suppressed.

\chapter{The Entropy Bound}

As discussed in \secref{wavenumber-spacing}, we
are interested in vacuum-bounded states for quite large energies.
When $E_0$ is large many different modes contribute to the entropy.
To get an accurate result for a system in which many modes are
important requires using many oscillators, which is
computationally intractable.  Instead, we take the even wavenumber
spacing of \secref{wavenumber-spacing} as given and derive
a bound on the entropy from that ansatz.

\section{The first outside oscillator}\label{sec:derive-bound}

With the wavenumber spacing fixed we have one free parameter, the
spacing $k_1$, which depends on the energy $E_0$.  To fix $k_1$ we we
examine the expectation value of $w_\Ninn^2$.  Since this is an
outside operator, it must have the same value in the vacuum as in a
vacuum-bounded state.  The vacuum value is straightforward to compute,
and we show below that in the vacuum-bounded state the value depends
only on the wavenumbers.  In fact, we will be able to derive only an
upper bound in the vacuum case and only a lower bound in the case of a
vacuum-bounded state.  However, these bounds are sufficient to derive
a lower bound on $k_1$ and thus an upper bound on $S(E)$.

\subsection{The vacuum}\label{sec:vacuum-calc}

In \secref{xopexform} we computed the values of $\xopex$.  To
convert to $\bw$ coordinates we proceed as follows:  From 
\secref{xfromw}, $\bx = Q \bw$ and so
\be[xfromw]
\bx_\out = R \bw_\out = D \bw_\mid + Y_\outgs \bw_\gs\,.
\ee
In the numerical work we found that with our choice for $D$ we got
\be
\tilde K_\inmid = \lowerleftnonzero{-g}\,.
\ee
Since
\be
\tilde K_\inmid = K_\inout D = \lowerleftnonzero{-g} D
\ee
it follows that the first row of $D$ is $(1,0\ldots 0)$.  Thus from 
\eqref{xfromw} we get
\be
x_\Ninn = w_\Ninn + (Y_\outgs)_1 \cdot \bw_\gs
\ee
where $(Y_\outgs)_1$ denotes the first row of $Y_\outgs$.
Now $\widetilde\wopsub_\gsgs = (1/2) I$ and $\widetilde\wopsub_\gsfree = 0$,
so
\be
\<x_\Ninn^2\> =  \<w_\Ninn^2\> + \half\left(Y_\gsout Y_\gsout^T\right)_{11}\,.
\ee
Since the last term is non-negative, we have
\be
\<w_\Ninn^2\>^\vac \le \<x_\Ninn^2\>^\vac\,.
\ee

From \eqref{finalxop} we have
\be
\<x_\Ninn^2\>^\vac = \xopexsub_{\Ninn,\Ninn}
 = {L_1 \over 2\pi} \left[\psi\left(2\Nin + {5\over 2}\right)
- \psi\left(\half\right)\right]
\ee
We would like to evaluate this expression in the
$\Nin\rightarrow\infty$ limit with $\Lin$ fixed.  There is a prefactor
of $L_1$, which goes to zero in this limit, but that is just an
artifact of the conventions we have used for the discrete problem, and
will appear in the finite-energy vacuum-bounded states as well.  For
large $x$,
\be[digamma-big]
\psi(x) \sim \ln x + O(1/x)\,,
\ee
so without the prefactor there is a logarithmic divergence.  We are
interested in the $\ln \Nin$ term, and in the constant term, but we
will ignore any terms of order $1/\Nin$ or lower.

We use \eqref{digamma-big} and $\psi(1/2) = -\gamma - 2\ln 2$, where
$\gamma$\notationlabel{gamma} is Euler's constant, to get
\be
\<x_\Ninn^2\>^\vac = {L_1\over 2\pi}\left[\ln 2\Nin + \gamma + 2\ln 2
+ O\left({1\over \Nin}\right)\right]
\ee
and so
\be[wwvacfinal]
\<w_\Ninn^2\>^\vac \le \<x_\Ninn^2\>^\vac
 = {L_1 \over 2\pi}\left[\ln 8\Nin+\gamma+O{\left(1\over \Nin\right)}\right]\,.
\ee

\subsection{Evenly spaced wavenumbers}

Now we will compute the same correlator in the vacuum-bounded system,
using the ansatz that the wavenumbers are multiples of some spacing
$k_1$.  From \eqref{wwexptval} we have
\be[wninndef]
\<w_\Ninn^2\> = \sum_\alpha {1\over 2\omega_\alpha}(w^\alpha_\Ninn)^2
\coth\left({\beta\over 2}\omega_\alpha\right)\,.
\ee
There are $\Nnorm\approx \pi/k_1$ normal modes which are sine waves in
the inside region.  These modes are equivalent to the modes that we
would have for a problem with a rigid boundary at position
\be
\notationlabel{Lin'}
\Linp \equiv \Nnorm L_1 \approx {\pi\over k_1} L_1 \,.
\ee
There are also $\Nfree-\Nnorm$ abnormal modes.  We do not know how to
compute the contribution of the the abnormal modes to the correlator.
However, the contribution from each mode is positive, so by taking
only normal modes we will find a lower bound on $\<w_\Ninn^2\>$.

\vspace{2pt}			
From \eqref{definewin} for the normal modes we have
\be[definexin]
w^n_\mu = N_n \sin \mu k_n
\ee
for $\mu=1\ldots\Ninn$\,.  Putting this in \eqref{wninndef} we get
\bea[outoutdef]
\<w_\Ninn^2\> & = &
\sum_{n=1}^{\Nnorm}
{1\over 2\omega_n}\coth\left({\beta\over 2}\omega_n\right)
{N_n}^2 \sin^2 k_n(\Ninn)+\text{abnormal modes}\nonumber\\
& \ge &
\sum_{n=1}^{\Nnorm}
{1\over 2\omega_n}\coth\left({\beta\over 2}\omega_n\right)
{N_n}^2 \sin^2 k_n(\Ninn)
\eea
since the abnormal mode contribution is positive.

The important point here is that we know $w^n_\mu$ for $\mu$ up
to $\Ninn$, and we know $\<w_\mu w_\nu\>$ for $\mu$ and $\nu$ down to
$\Ninn$.  Thus taking $\mu=\nu=\Ninn$ gives the unique correlator for
which we know the components that go into the expression for the
correlator while also knowing that the correlator must have the same
value as in the vacuum.  The same argument does not work for
$\<P_{w_\Ninn} P_{w_\Ninn}\>$, because we know the inverse mode matrix only up
to $\mu = \Nin$ and not $\mu = \Ninn$.

Now we will compute the right-hand side of \eqref{outoutdef} in the limit
where $\Nin$ and $\Nnorm$ are large.  We will ignore all terms of
$O(1/\Nin)$ or $O(1/\Nnorm)$ and so take $\Linp/\Nnorm = \Lin/\Nin$.
We use
\be
k_n = n k_1 = {\pi n\over \Nnorm}
\ee
and
\be[defvbfreqs]
\omega_n = {2\over L_1}\sin{k_n\over 2} ={2\over L_1}\sin{\pi n\over 2\Nnorm}\,.
\ee

First we must derive the normalizations of the normal modes.  With our
choice of normalization,
\be
\sum_n w^n_\mu w^n_\nu = T'_{\mu\nu}\,.
\ee
Thus if $a$ is an inside oscillator, then for any $\mu$,
\be
\sum_n w^n_a w^n_\mu = \delta_{a\mu}
\ee
so 
\be[orthogonality]
\delta_{a\mu}  = \sum_{n=1}^{\Nnorm} N_n^2
\sin k_n a\sin k_n \mu + \text{abnormal modes}\,.
\ee
We would like to use \eqref{orthogonality} to determine the
normalization factors $N_n$, but first we have to dispose of the
abnormal mode term.  Later we will find that $\Nnorm$ is close to
$\Nin$.  Using this, and since we are working in the limit where
$\Nin$ is large, we can choose $a$ such that
\be[abounds]
1 \ll a \ll 2\Nin - \Nnorm \sim \Nin\,.
\ee
Each abnormal mode $n$ contributes 
\be
N_n^2 \sinh k_n a \sinh k_n \mu \equiv \deltaabnn
\ee
to the right-hand side of \eqref{orthogonality}.  However, this
same mode contributes
\be
{L_1\over 4} N_n^2 {\sinh^2 k_n (\Ninn) \over \cosh (k_n/2)}
\ee
to the sum for $\<w_\Ninn^2\>$ in \eqref{outoutdef}.
Since $\<w_\Ninn^2\> \le L_1/(2\pi)[\ln\Nin + O(1)]$ it follows that
\be
N_n^2 {\sinh^2 k_n (\Ninn) \over \cosh (k_n/2)} \lesssim {2\over \pi}\ln \Nin
\ee
for each $n$ and thus that
\be
\deltaabnn < {2\sinh k_n a\sinh k_n \mu \cosh(k_n/2)\over
	\pi \sinh^2 k_n(\Ninn)} \ln \Nin\,.
\ee

The exact values of the $k_n$ for the abnormal modes vary with
$\Nin$.  For successive $\Nin$ values, each $k_n$ decreases toward $0$
until the corresponding mode converts to a normal mode as described in
\secref{modeconvert}.  By avoiding the points where these
``conversions'' are about to take place it is possible to find a
sequence of values for $\Nin$ which have all $k_n \gtrsim 1$.  For such
values,
\be[deltaabnsmall]
\deltaabnn \lesssim e^{-k_n\left(2 \Nin -a-\mu+3/2\right)}\,.
\ee
From \eqref{abounds} the exponent in \eqref{deltaabnsmall} is
$\ll -1$, so $\deltaabnn$ is exponentially small.  Since
there are at most $\Nin$ abnormal modes, their total contribution
to (\ref{eqn:orthogonality}) is exponentially suppressed.

Thus we ignore the abnormal modes in \eqref{orthogonality},
multiply by $\sin k_m \mu$ and sum over $\mu$ to get
\bea
\sin k_m a & = & \sum_{\mu=1}^{\Nnorm}\sin k_m\mu
\sum_{n=1}^{\Nnorm} N_n^2\sin k_n a\sin k_n \mu \nonumber\\
& = & \sum_{n=1}^{\Nnorm} N_n^2 \sin k_n a \sum_{\mu=1}^{\Nnorm}
  \sin\text{$\pi m \mu\over \Nnorm$}
  \sin \text{$\pi n \mu\over \Nnorm$} \nonumber\\
& = &\sum_{n=1}^{\Nnorm}N_n^2 \sin k_n a \cdot {\Nnorm\over 2}
 \delta_{mn}\nonumber\\
& = & N_m^2 {\Nnorm\over 2} \sin k_m a\,,
\eea
from which we conclude that
\be
N_m^2 = {2\over\Nnorm}
\ee
for each mode $m$.
Putting this in \eqref{outoutdef} gives
\be[woutoutprelim]
\<w_\Ninn^2\> \gtrsim \sum_{n=1}^{\Nnorm} 
{\sin^2 k_n(\Ninn) \over\Nnorm \omega_n}
\coth\left({\beta\over 2}\omega_n\right)\,.
\ee
Note that \eqref{woutoutprelim} does not depend on $N$ but only
$\Nnorm$.

Let us define a dimensionless parameterization of the departure from
the vacuum state,
\blea
\notationlabel{Delta}\notationlabel{tau'}
\Delta &=& {\Nnorm - \Nin\over\Nnorm} \approx {\Linp - \Lin\over\Linp}
  \label{eqn:Deltadefined}\\
\tau' &=& \Linp T = \Linp/\beta\label{eqn:taupdefined}\,.
\notationlabel{T}
\elea
Then $\sin^2 k_n (\Ninn) = \sin^2 n\pi (1-\Delta)
 = \sin^2 n\pi\Delta$.
With $\omega_n$ from \eqref{defvbfreqs} we get
\be
\<w_\Ninn^2\> \gtrsim {L_1\over 2}
\sum_{n=1}^{\Nnorm}
   {\sin^2 n\pi\Delta \over\Nnorm \sin {n\pi\over 2\Nnorm}}
\coth\left({\beta\over L_1}\sin{n\pi\over 2\Nnorm}\right)\,.
\ee

In the limit of large $\Nfree$, the argument of $\coth$ becomes $\beta
\pi n / (2 L_1 \Nnorm) \approx \pi n/(2 \tau')$.  We  expand
$\coth x = 1+2/(e^{2x} -1)$ to get
\be[bound1]
\<w_\Ninn^2\> \gtrsim {L_1\over 2}
\sum_{n=1}^{\Nnorm}
   {\sin^2 n\pi\Delta \over\Nnorm \sin {n\pi\over 2\Nnorm}}
\left(1+{2\over(e^{-\pi n/\tau'}-1)}\right)\,.
\ee
We work first with the term not involving $\tau'$.  We expand the
numerator using $\sin^2 x = (1-\cos 2x)/2$ to get
\be[boundexp]
{L_1\over 4}\sum_{n=1}^{\Nnorm} \left({1 \over \Nnorm\sin {n\pi\over 2\Nnorm}}
   - {\cos 2n\pi\Delta \over \Nnorm\sin {n\pi\over 2\Nnorm}}\right)\,.
\ee
This first term is the one that diverges as $\Nnorm\rightarrow\infty$.
We can separate out the divergent part to get
\be[bounddiverg]
{L_1\over2\pi}\sum_{n=1}^{\Nnorm} {1\over n} 
+ {L_1\over 4}\sum_{n=1}^{\Nnorm} \left({1\over \Nnorm\sin 
{n\pi\over 2\Nnorm}} - {2\over\pi n} 
\right)\,.
\ee
The first term of \eqref{bounddiverg} gives
\be[bound-term111]
{L_1\over2\pi}\left(\ln\Nnorm + \gamma\right)\,.
\ee
The second term of \eqref{bounddiverg} is finite and can be converted
to an integral in the $\Nnorm\rightarrow\infty$ limit, to give
\be[bound-term112]
{L_1\over4}\int_0^1 dx\left({1\over\sin {\pi x\over 2}}-{2\over \pi x}\right)
= {L_1\over2\pi}\left[\ln\left({1\over x}\tan{\pi x\over 4}\right)\right]^1_0\\
= {L_1\over2\pi}\ln{4\over \pi}\,.
\ee
The remaining term of \eqref{boundexp} is
\be
-{L_1\over 4}\sum_{n=1}^{\Nnorm} {\cos 2 n\pi\Delta\over \Nnorm \sin{n\pi\over 2 \Nnorm}}\,.
\ee
To compute this we use
$1/\sin x = \csc x = 1/x + x/6 + 7x^3/360 + \cdots\,$ to get
\be
-{L_1\over 4}\sum_{n=1}^{\Nnorm}
\left({2\over\pi n} + {\pi\over 12}{n\over \Nnorm^2}
+{7\pi\over 2880}{n^3\over \Nnorm^4} + \cdots\right)\cos 2 n\pi \Delta\,.
\ee
The first term can be summed in the $\Nnorm\rightarrow\infty$ limit,
\be[bound-term12]
- {1\over2\pi}\sum_1^\infty{\cos 2 n\pi\Delta\over n}
= {1\over2\pi}\ln \left(2 \sin \pi\Delta\right)\,.
\ee
The rest of the terms do not contribute.  Because of the
oscillations of the cosine, $\sum_1^{\Nnorm} n^k \cos n\pi\Delta$ goes as
$\Nnorm^k$ rather than $\Nnorm^{k+1}$ and thus is killed by the corresponding
$\Nnorm^{k+1}$ in the denominator.

Putting \eqsref{bound-term111}, (\ref{eqn:bound-term112}) and
(\ref{eqn:bound-term12}) together, the first term on the right of
\eqref{bound1} gives
\be[bound-term1]
{L_1\over2 \pi}\left(\ln{8\Nnorm\sin\pi\Delta\over\pi}+\gamma\right)\,.
\ee

We now look at the second term of \eqref{bound1},
\be
L_1
\sum_{n=1}^{\Nnorm} {\sin^2 n\pi\Delta \over\Nnorm \sin {n\pi\over 2\Nnorm}}
 {1\over(e^{-\pi n/\tau')}-1)}\,.
\ee
Here the divergence is cut off by the exponential in the denominator.
We again expand using $1/\sin x = 1/x + x/6 + \cdots$.  The first term has
no $\Nnorm$ dependence and we can extend the sum to $\infty$.  In
the next term, the sum is cut off by the exponential in the
denominator, leading to a term of order $(\tau'/\Nnorm)^2$.  Further
terms have higher powers of $\tau'/\Nnorm$.  In the limit
$\Nnorm\rightarrow\infty$ we ignore all these terms, which leaves
\be[boundsimpsum]
{2L_1\over\pi}\sum_{n=1}^\infty {\sin^2 n\pi\Delta \over
n \left(e^{\pi n/\tau'}-1\right)}\,.
\ee
We are interested in the high-energy limit, for which $\tau' \gg 1$.
Later we will see that $\Delta$ is of order $\ln\tau'/\tau' \ll 1$.
Thus the summand in \eqref{boundsimpsum} is slowly varying and we
can convert the sum into an integral,
\be[boundint]
{2L_1\over\pi}\int_0^\infty {\sin^2 \pi\Delta x\, dx\over
x\left(e^{\pi x/\tau'}-1\right)}\,.
\ee
The error in \eqref{boundint} is approximately the term that we
would have for $n\rightarrow 0$ in \eqref{boundsimpsum}.  Taking this
limit we find that the error has order $\Delta^2\tau'
\sim (\ln\tau')^2/\tau' \ll 1$, so our approximation is good.
The integral in \eqref{boundint} can be done and the result
is
\be[bound-term2]
L_1 \tau'\Delta
   + {L_1\over2\pi}\ln{1-e^{-4 \pi\tau' \Delta}\over 4 \pi\tau' \Delta}\,.
\ee
Putting together \eqsref{bound-term1} and (\ref{eqn:bound-term2}) we find
\be
\<w_\Ninn^2\> \gtrsim 
 {L_1\over2 \pi}\left(\ln{8\Nnorm\sin\pi\Delta\over\pi}+\gamma
+2\pi\tau'\Delta+\ln{1-e^{-4 \pi\tau' \Delta}\over 4 \pi\tau' \Delta}
\right)\,.
\ee
 
Now we set $\<w_\Ninn^2\>$ = $\<w_\Ninn^2\>^\vac$ from \eqref{wwvacfinal}
to get
\be
\ln {8\Nin\over\pi} \gtrsim \ln{8\Nnorm\sin\pi\Delta\over\pi}
+2\pi\tau'\Delta+\ln{1-e^{-4 \pi\tau' \Delta}\over 4 \pi\tau' \Delta}
\ee
or
\be
2\pi\tau'\Delta+\ln{1-e^{-4 \pi\tau' \Delta}\over 4 \pi\tau' \Delta}
\alt \ln{\Nin\over\Nnorm\sin\pi\Delta}\,.
\ee
Now we use $\Nin/\Nnorm = 1-\Delta$ from \eqref{Deltadefined} and
approximate $\sin\pi\Delta \approx \pi\Delta$ since $\Delta$ is
small.  Since this is already $O(\Delta)$ we then approximate
$(1-\Delta)/\Delta \approx 1/\Delta$ to get
\be
2\pi\tau'\Delta+\ln{1-e^{-4 \pi\tau' \Delta}\over 4 \pi\tau' \Delta}
\alt \ln{1\over\pi\Delta}\,.
\ee
Thus $\Delta \le \Delta_{\max}$ where
\be[deltamax]
\Delta_{\max}
 = {1\over 2 \pi\tau'}\ln{4\tau'\over 1-e^{-4\pi\tau'\Delta_{\max}}}\,.
\ee
If instead of $\tau' = \Linp T$ we use
\be
\notationlabel{tau}
\tau \equiv \Lin T
\ee
we will make an
error of order $\Delta_{\max}$, which we expect to be small.  We ignore
this second order contribution and take $\tau'$ as $\tau$ in \eqref{deltamax},
\be
\Delta_{\max}
 = {1\over 2 \pi\tau}\ln{4\tau\over 1-e^{-4\pi\tau\Delta_{\max}}}\,.
\ee
If we ignore $e^{-4\pi\tau\Delta_{\max}}$ in the denominator we get
\be
\Delta_{\max} = {1\over 2\pi\tau}\ln{4\tau}\,.
\ee
Using this we find that 
$e^{-4\pi\tau\Delta_{\max}} = (4\tau)^{-2} \ll 1$
since $\tau \gg 1$, which justifies ignoring this term.  We will also
ignore $\ln 4$ by comparison with $\ln \tau$.
Thus we conclude
\be
\Delta \lesssim {1\over 2\pi\tau}\ln \tau  + O\left({1\over\tau}\right)
= {1\over 2\pi\Lin T}\ln{\Lin T} + O\left({1\over\Lin T}\right)
\ee
and
\be
\Linp \le \Lin + {1\over 2\pi T} \ln {\Lin T} +O\left({1\over T}\right)\,.
\ee
The equivalent system is larger by at most a thermal wavelength times a
logarithmic factor depending on the inside size.

\section{Propagation of bounds}\label{sec:bound-prop}

In the previous section we derived an expression that gives the
frequencies, and thus the entropy, for a vacuum-bounded system at a
given temperature $T = 1/\beta$.  Given such an expression, we would
like to compute the entropy as a function of energy.  Unfortunately
the energy is not simple to compute from the frequencies
alone.\footnote{Such a computation can be done, but since $E$ needs to
be renormalized against the ground-state energy of the entire system,
the result depends sensitively on the frequencies and normalizations
even for very high-energy modes.}  However, we can easily compare the
entropy of the vacuum-bounded system to that of a system with a rigid
boundary at $\Lin$ and the same temperature.  To make this comparison
at fixed energy instead, we proceed as follows.

Consider the free energy $F=E-TS$ which has $dF= -S dT$.  Integrating
gives
\be[fdef]
E-TS= -\int_0^T S(T') dT'\,.
\ee
Let $S^\rb$ and $T^\rb$ denote the entropy and temperature of the system
with a rigid boundary at $\Lin$, and $S$ and $T$ denote those
for the vacuum bounded system. For any quantity $A$ let $\ddelta
A(T)$\notationlabel{d(T)} 
denote the difference between vacuum-bounded and rigid box systems
at fixed temperature, $\ddelta A(T) \equiv A(T) - A^\rb(T)$, and $\ddelta
A(E)$\notationlabel{d(E)}
denote the same difference at fixed energy, $\ddelta A(E) \equiv
A(E) - A^\rb(E)$.  With $E$ fixed we compare the differences (to first
order) in the two sides of \eqref{fdef} between the
vacuum-bounded and rigid box systems,
\be
-T \ddelta S(E) - \ddelta T(E) S = - \ddelta T(E) S - \int_0^T \ddelta
S(T') dT'\,,
\ee
where the first term on the right-hand side comes from the change in
the integration limit.  Thus
\be[bound-convert]
\ddelta S(E) = {1\over T} \int_0^T \ddelta S(T') dT'\,.
\ee

\section{The final entropy bound}

Now we apply \eqref{bound-convert} to the case of \secref{derive-bound}
where
\be
\omega_n \approx {n\pi\over \Linp}
\qquad\text{and}\qquad
\Linp \approx \Lin(1+\Delta)
\ee
with
\be
\Delta \Lin \le {1\over 2\pi T}\ln \Lin T\,.
\ee
At any given temperature, the vacuum-bounded system has the entropy
$S(T)$ of a system of length $\Linp$.  Now in a one-dimensional system
the entropy density is proportional to the temperature,
\be
S^\rb =  {\pi\over 3}\Lin T\,,
\ee
and thus the entropy difference between vacuum-bounded and
rigid box systems is
\be
\ddelta S(T) = {\pi\over 3}\Delta\Lin T \le {1\over 6}\ln\Lin T\,.
\ee
Using \eqref{bound-convert} we get\footnote{It happens that
$\ddelta S(E)$ and $\ddelta S(T)$ are approximately the 
same, but that is a particular property of the system at hand.  For
example, if $\Delta$ were a constant we would have
$\ddelta S(E) = 1/T\int_0^Tc\Delta\Lin T' dT' = 2 c\Delta\Lin T
= 2 \ddelta S(T).$}
\be
\ddelta S(E) \le {1\over T}\int_0^T {1\over 6}\ln\Lin T' dT'
= {1\over 6}\left(\ln\Lin T-1\right)\,.
\ee
Since we are ignoring terms of order $1$ by comparison with those
of order $\ln T$, we can write
\be
\ddelta S(E) \lesssim {1\over 6}\ln\Lin T \approx {1\over 6}\ln S^\rb\,.
\ee

Thus we conclude that the vacuum-bounded condition closely approximates
the rigid box of length $\Lin$.  For the same energy, the
vacuum-bounded condition allows slightly more entropy.  The entropy
difference grows at most logarithmically with rigid box entropy.
For high energies, $S^\rb \gg 1$ so we conclude that $\ddelta S
\ll S^\rb$.

\section{Discussion}

We have introduced a new way of specifying that matter and energy are
confined to a particular region of space.  Rather than giving a
boundary condition per se, we specify a condition on a density matrix
describing the state of the overall system.  We require that any
measurement which does not look into the inside region cannot
distinguish our system from the vacuum.  This avoids certain
difficulties such as the Casimir energy that results from the
introduction of a boundary and the geometric entropy
\cite{srednicki:geom-ent,callan:geom-ent} that results from ignoring
part of a system.  For these {\em vacuum-bounded} states, we consider
the problem of finding the maximum-entropy state for a given total
energy.  This is analogous to the problem of finding the thermal
state in a system with a rigid boundary.

Unfortunately, the vacuum-bounded problem is more difficult than the
analogous problem with a rigid boundary and we must resort to working
in one dimension and to numerical solution on a lattice.  It is,
however, possible to reduce the problem to a finite number of degrees
of freedom, even when the outside region is infinitely large.  From
the numerical solution we justify the ansatz that the continuum
wavenumbers are evenly spaced in this problem.  Using this ansatz we
compute an upper bound on the entropy of a vacuum-bounded state, and
show that for high energies ($ER \gg 1$) the entropy approaches that
of a system with rigid boundaries.  Of course this is what one would
expect for a system whose typical wavelengths are much shorter than
the size of the inside region.

To apply this result to an evaporating black hole we look at the state
produced by the black hole after evaporation \cite{preskill:review}.
Since our calculation was one-dimensional we must assume that the
similarity between the vacuum-bounded state and the thermal state with
a rigid boundary extends to three dimensions.  Then we infer that very
little entropy can be emitted in the final explosion, in accord with
the results of Aharonov, Casher and Nussinov \cite{aharonov:orig} and
Preskill \cite{preskill:review}.  For example, a black hole formed in
the big bang with mass of order $10^{15}g$ would be evaporating today.
During its life it would have radiated entropy $S\sim 10^{38}$.  Now
we assume that the entropy of the final explosion has energy $E\sim
10^{19}\text{erg}$ contained in radius $R\sim 10^{-23}\text{cm}$ as in
\secref{realistic-er}, and that the maximum entropy is not too
different from that of a spherical box, in accord with our
one-dimensional result.  Then we find that the final explosion can emit
only entropy $S \sim 10^{10}$, which is a factor of $10^{28}$ less
than what was emitted earlier in the thermal radiation.  The choice of
$\Tunk$ is somewhat arbitrary, but whatever value one chooses there is
some fixed bound on the emission of entropy after $\Tunk$ is reached.
By considering a sufficiently large starting black hole, and thus
sufficient entropy emission at early times, one always finds that the
late time information is too little to produce a final pure state.

This argument means that a black hole must not evaporate completely
but rather leave a remnant or remnants, that information must be lost,
or else that the Hawking radiation is not exactly thermal, even at
very early times \cite{esko:nosemiclass}.

\appendix
\makeatletter
\clearpage
\thispagestyle{plain}		
\cleardoublepage
\thispagestyle{plain}
\global\@topnum\z@
\@afterindentfalse
\refstepcounter{chapter}
\addcontentsline{toc}{chapter}{Appendix \hspace{0.5em}Low Energy Results}
\markboth{Appendix}{}
\markright{A. Low Energy Results}
\vspace*{50pt}
{\parindent 0pt \raggedright 
 \Huge\bf Appendix\par
 \vskip 20pt Low Energy Results
 \nobreak \vskip 40pt }
\makeatother

In the low-energy regime we do not see the linear wavenumber relation
that we see in the high-energy case (\figref{fig4-9}).
Instead, for sufficiently low temperature, the wavenumbers and
frequencies are nearly the same as in the reduced vacuum.  These
frequencies are much lower than in a rigid box with the same $\Lin$.
At low temperatures, the entropy depends only on the low-lying
frequencies and on $\beta$.  Thus we expect that there will be
significantly more entropy in a vacuum-bounded state than a rigid box
state of the same energy.  While we don't know how to construct an
analytic proof of this claim, we will outline a general argument here,
and make a conjecture supported by numerical data.

In the low-energy regime we can make a first-order expansion around
the vacuum.  To do this we note that the only dependence on $\beta$ in
our equations is through $\coth(\beta\omega_\alpha/2)$ in \eqsref{wexptvals}.
For large $\beta$ we can approximate
\be
\coth{\beta\omega_\alpha\over 2} \approx 1+2e^{-\beta\omega_\alpha}\,.
\ee
The change in $\coth(\beta\omega_\alpha/2)$ is the largest for the
smallest frequency, which we will call $\omega_1$.  We will ignore
$e^{-\beta\omega_\alpha}$ for larger $\omega_\alpha$ by comparison
with $e^{-\beta\omega_1}$.  Thus we take
\blea[loweapproxfirst]
\ddelta\coth{\beta\omega_1\over2} &=& 
2e^{-\beta\omega^\vac_1} \equiv 2\epsilon \\
\ddelta\coth{\beta\omega_\alpha\over 2} &=& 0 \qquad\hbox{for $\alpha> 1$}\,.
\elea
Then we write
\blea
T'_\midmid &=& \tilde T_\midmid + \ddelta T_\midmid\\
K'_\midmid &=& \tilde K_\midmid + \ddelta K_\midmid
\elea
where $\ddelta K_\midmid$ and $\ddelta T_\midmid$ are $O(\epsilon)$.
These changes give rise to $O(\epsilon)$ changes in $U$ and
the $\omega_\alpha$, which in turn give rise to $O(\epsilon)$ changes in
$\<w_m w_n\>$ and $\<P_{w_m} P_{w_n}\>$.
\abovedisplayskip 10pt plus3pt minus5pt
\belowdisplayskip \abovedisplayskip

Since overall $\<w_m w_n\>$ and $\<P_{w_m} P_{w_n}\>$ cannot change we
must have
\blea[loweapproxlast]
0 = \ddelta\<w_m w_n\> = \sum_\alpha\bigg(&&
-{\ddelta\omega_\alpha\over 2\omega_\alpha^2} U^\vac_{m\alpha} U^\vac_{n\alpha}
+ {1\over 2\omega^\vac_\alpha} \ddelta U_{m\alpha} U^\vac_{n\alpha}
+ {1\over 2\omega^\vac_\alpha} U_{m\alpha} \ddelta U_{n\alpha}\bigg) \nonumber\\
&&+ {1\over \omega^\vac_1} U^\vac_{m1} U^\vac_{n1} \epsilon\\
0 = \ddelta\<P_{w_m} P_{w_n}\> = \sum_\alpha\bigg(&&
{\ddelta\omega_\alpha\over 2}\uvacinv_{\alpha m} \uvacinv_{\alpha n} 
+ {\omega^\vac_\alpha\over 2}\ddelta U^{-1}_{\alpha m} \uvacinv_{\alpha n}
+{\omega^\vac_\alpha\over 2}\uvacinv_{\alpha m}\ddelta U^{-1}_{\alpha n}\bigg)
\nonumber\\
&&+ \omega^\vac_1 \uvacinv_{m1} \uvacinv_{n1}\epsilon\,.
\elea

We thus have $\Nin(\Nin+1)$ linear equations for $\Nin(\Nin+1)$
unknown values of $\ddelta T_\midmid$ and $\ddelta K_\midmid$, which are
readily solved.  Since the inhomogeneous part of these equations is
$O(\epsilon)$, all the results must be $O(\epsilon)$ as well.  In
particular, the $\ddelta\omega_\alpha$ are $O(\epsilon)$.  Now
if $T$ is very small as compared to all the $\omega_\alpha$, then
$\epsilon$ will be small as compared to all the parameters of the
problem, and so the first-order approximation will be good.  For any
fixed number of oscillators $\Nin$ there will be some minimum
frequency $\omega_1$, and if we take $\beta\ll 1/\omega_1$ we will
always be in this regime.

Now the entropy $S$ depends only on $\beta$ and the $\omega_\alpha$.
Since the modes are uncoupled,
\be
S = \sum_\alpha S_1(\beta\omega_\alpha)
\ee
with
\be
S_1(\beta\omega) = -\ln(1-e^{-\beta\omega})
 + {\beta\omega\over e^{-\beta\omega} -1}\,.
\ee
Since $\beta\omega_\alpha \gg 1$ all the terms are very small, and the
$\omega_1$ term dominates,
\be
S \approx S_1(\omega_1) = (1+\beta\omega_1)e^{-\beta\omega_1} +
O(e^{-2\beta\omega_1})\,.
\ee
Since $\epsilon$ drops exponentially with increasing $\beta$, we
expect that for $\beta$ large enough, $\beta\ddelta\omega_1 \ll 1$ so
that
\be[loweS]
S = (1+\beta\omega^\vac_1)e^{-\beta\omega^\vac_1}+O(\epsilon^2)\,.
\ee
The value of $S$ given in \eqref{loweS} is the one we would get
from a rigid box with length
\be[lowelinprimedef]
\Linp = \pi/\omega^\vac_1\,.
\ee

To approximate the energy, we proceed along the lines of 
\secref{bound-prop}.  The direct calculation is made difficult by the
fact that, while $H'$ differs from $H$ only by $O(\epsilon)$, we must
subtract from both Hamiltonians a large ground-state energy.  Instead we
work by integrating on $T$.  From \eqref{fdef} we have
\be[loweint]
E(T) = TS(T) - \int_0^T S(T') dT'\,.
\ee
Now $\Linp$ depends only on $\omega_1^\vac$, which depends on $\Nin$
but not on $\beta$.  If \eqref{loweS} is valid for a particular $\Nin$
at $\beta=1/T$ is it valid for $T'<T$ and $\beta'=1/T'>\beta$.  
Thus both $S(T)$ and $S(T')$ in \eqref{loweint} are just the entropy of a
rigid box of length $\Linp$.  Thus the entropy-to-energy
relationship is just $S(E) = S^\rb(\Linp;E)$, the entropy as a
function of the given energy in a rigid box with length
$\Linp$.

For such a rigid box at very low energy we find
\be
E={\omega^\rb_1\over e^{\beta\omega^\rb_1} -1}\approx \omega^\rb_1
e^{-\beta\omega^\rb_1}
\ee
and thus
\be
S = \left(1+\ln{\omega^\rb_1\over E}\right){E\over\omega^\rb_1}
\ee
where $\omega^\rb_1 = \pi/\Linp = \omega^\vac_1$ is the frequency of
the lowest mode.

Now for any given $\Nin$ we get some $\omega^\vac_1$.  As discussed in
\secref{smallfreqs}, the larger $\Nin$ we choose, the smaller
$\omega^\vac_1$ we will have.  For $\Nin$ fixed we can choose
$\beta\gg1/\omega^\vac$ and proceed as above to get a large value of
$\Linp$.  However, we are really interested in the continuum limit at
fixed temperature.  If we increase $\Nin$ with $\beta$ fixed we will
find that $\omega_1$ (and eventually an arbitrary number of the
$\omega_\alpha$) will become smaller than $1/\beta$.  When this
happens, the approximations of Eqs.\
(\ref{eqn:loweapproxfirst}--\ref{eqn:loweapproxlast}) will no longer
be good.

However, we do not expect the entropy to decrease drastically in this
limit.  To make the entropy small would require making all the
frequencies large.  If the frequencies were large, the approximations
we have used would again become valid.  Then we could argue as before
that the entropy should be large.  It would be hard to have a
consistent picture.

Now consider the limit as $T\rightarrow 0$.  For each $T$ we start
with some initial number of oscillators $\Nin^{(0)}$.  We choose
$\Nin^{(0)}$ not too large, such that $\omega^\vac_1 \gg T$.  With
this value of $\Nin$, we find $\Linpsup{0} \sim \pi/\omega^\vac_1$.
We then let $\Nin\rightarrow\infty$ and we conjecture that the entropy
does not change much, and thus in the continuum limit $S(E) \sim
S^\rb(\Linpsup{0}; E)$.  As we decrease $T$ we can decrease the
initial $\omega^\vac_1$ and so increase $\Linpsup{0}$ without bound.
Thus we make the following conjecture:
\begin{quotation}
For a given energy $E$, let $\Linp(E)$ be the length of a rigid
box such that the vacuum-bounded state with energy $E$ and length
$\Linp$ has entropy $S(E) = S^\rb(\Linp(E);E)$.  Then
\be
\lim_{E\rightarrow 0} {\Linp(E)\over \Lin} = \infty\,.
\ee
\end{quotation}

To support this conjecture numerically we turn to direct calculation
of energy and entropy values for vacuum-bounded states at low
temperature.  For various fixed values of $\beta=1/T$ and for various
numbers of oscillators we compute $S$ and $E$ and from them the
equivalent length $\Linp$. The results are plotted in
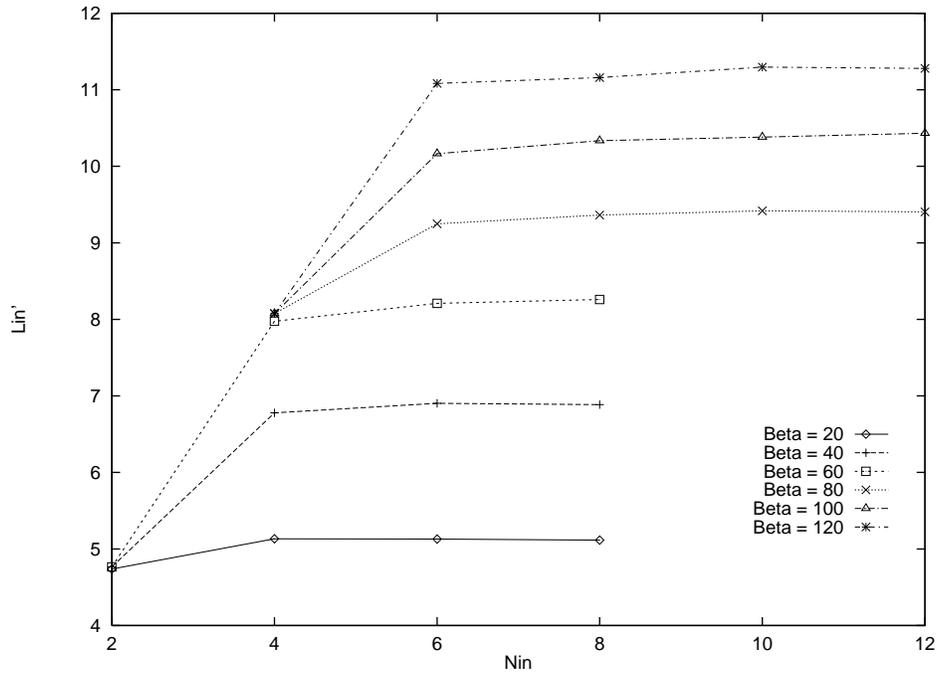
\mypsfigr{figa-1}{The length $\Linp$ of a rigid box that gives
the same $S(E)$ as a vacuum-bounded state at temperature $T=1/\beta$
and $\Lin = 1.0$}.  While $\Nin$ is still small enough for the
approximations Eqs.\
(\ref{eqn:loweapproxfirst}--\ref{eqn:loweapproxlast}) to be valid,
$\Linp$ grows with $\Nin$.  Once $\Nin$ has left this regime, it
appears that $\Linp$ levels off.  It is at least reasonable to believe
that there is no further change in $\Linp$ as $\Nin\rightarrow\infty$.
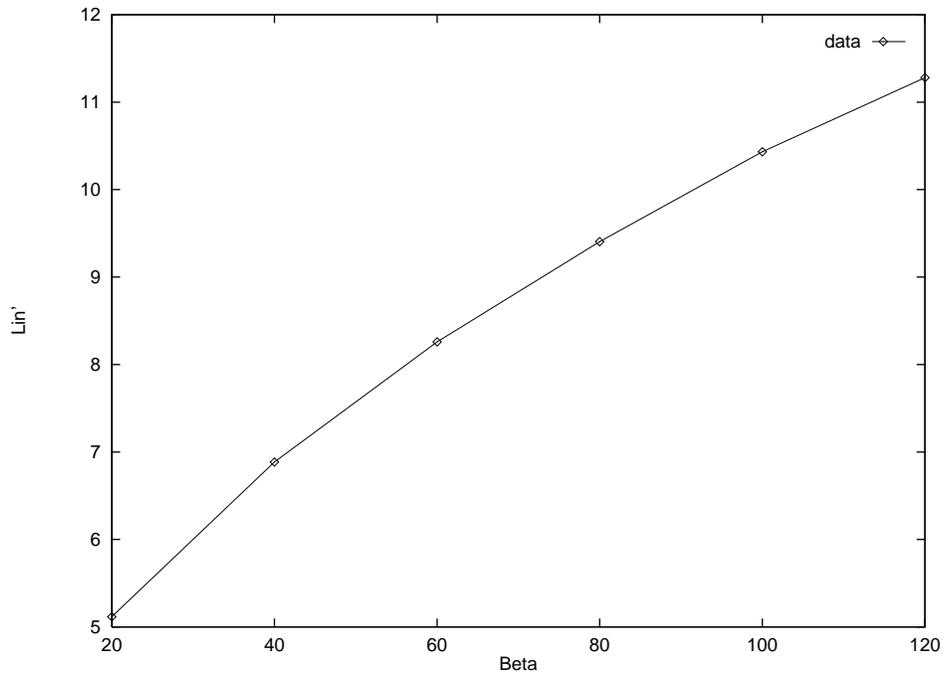
In \mypsfigr{figa-2}{The length $\Linp$ of a rigid box
with the same $S(E)$ plotted against $\beta$.  Each point is the value
for the largest number of oscillators available} we plot the eventual
level of $\Linp$ versus $\beta$.  It appears that the limiting value
of $\Linp$ grows nearly linearly with $\beta$, and thus
$\Linp\rightarrow\infty$ as $E\rightarrow 0$ as conjectured.



\end{document}